\documentclass[twocolumn]{aastex7}

\usepackage[usenames,dvipsnames]{xcolor}
\usepackage{physics}
\usepackage{amsmath}
\usepackage{CJK}

\usepackage{graphicx}
\usepackage{subfig}
\usepackage{multirow}

\usepackage{natbib}

\shorttitle{Untangling Similarities and Differences in SFHs Across Cosmological Simulations}
\shortauthors{Savelli et al.}

\newcommand{\MSun}{\mathrm{M}_\odot}
\newcommand{\SM}{M_\star}
\newcommand{\SFR}{\dot{M}_\star}
\newcommand{\sSFR}{\dot{M}_\star/m_\star}

\graphicspath{{./}{Figures_old}}

\begin{document}
\begin{CJK*}{UTF8}{gbsn}

\title{Untangling Similarities and Differences in Star Formation Histories Across Cosmological Simulations}

\author[0000-0003-0147-9562]{Alicia M. Savelli}
\affiliation{Canadian Institute for Theoretical Astrophysics, University of Toronto, 60 St. George Street, Toronto, ON M5S 3H8, Canada}
\affiliation{David A. Dunlap Department of Astronomy \& Astrophysics, University of Toronto, 50 St. George Street, Toronto, ON M5S 3H4, Canada}
\email{alicia.savelli@utoronto.ca}

\author[0009-0002-6356-3495]{Matthew A. Kustec}
\affiliation{David A. Dunlap Department of Astronomy \& Astrophysics, University of Toronto, 50 St. George Street, Toronto, ON M5S 3H4, Canada}
\affiliation{Department of Statistical Sciences, University of Toronto, 9th Floor, Ontario Power Building, 700 University Ave, Toronto, ON M5G 1Z5, Canada}
\email{matthew.kustec@mail.utoronto.ca}
\correspondingauthor{Alicia M. Savelli}

\author[0000-0003-2573-9832]{Joshua S. Speagle (沈佳士)}
\affiliation{David A. Dunlap Department of Astronomy \& Astrophysics, University of Toronto, 50 St. George Street, Toronto, ON M5S 3H4, Canada}
\affiliation{Department of Statistical Sciences, University of Toronto, 9th Floor, Ontario Power Building, 700 University Ave, Toronto, ON M5G 1Z5, Canada}
\affiliation{Dunlap Institute for Astronomy \& Astrophysics, University of Toronto, 50 St. George Street, Toronto, ON M5S 3H4, Canada}
\affiliation{Data Sciences Institute, University of Toronto, 17th Floor, Ontario Power Building, 700 University Ave, Toronto, ON M5G 1Z5, Canada}
\email{j.speagle@utoronto.ca}

\author[0000-0001-9298-3523]{Kartheik G. Iyer}
\affiliation{Institute for Astronomy, University of Hawaiʻi, 2680 Woodlawn Drive, Honolulu, HI 96822, USA}
\affiliation{Columbia Astrophysics Laboratory, Columbia University, 550 West 120th Street, New York, NY 10027, USA}
\affiliation{Dunlap Institute for Astronomy \& Astrophysics, University of Toronto, 50 St. George Street, Toronto, ON M5S 3H4, Canada}
\email{kiyer@hawaii.edu}

\author{Eshal Ahmed}
\affiliation{Dunlap Institute for Astronomy \& Astrophysics, University of Toronto, 50 St. George Street, Toronto, ON M5S 3H4, Canada}
\email{eshal.arshad@mail.utoronto.ca}

\author[0009-0003-0677-5675]{Zehao Peng (彭则皓)}
\affiliation{David A. Dunlap Department of Astronomy \& Astrophysics, University of Toronto, 50 St. George Street, Toronto, ON M5S 3H4, Canada}
\email{harry.peng@mail.utoronto.ca}

\author[0000-0002-8659-3729]{Norman Murray}
\affiliation{Canadian Institute for Theoretical Astrophysics, University of Toronto, 60 St. George Street, Toronto, ON M5S 3H8, Canada}
\affiliation{David A. Dunlap Department of Astronomy \& Astrophysics, University of Toronto, 50 St. George Street, Toronto, ON M5S 3H4, Canada}
\affiliation{Department of Physics, 60 St. George Street, Toronto, ON M5S 1A7, Canada}
\email{murray@cita.utoronto.ca}

\author[0009-0004-4042-3937]{Cissy Kuang (况诗其)}
\affiliation{David A. Dunlap Department of Astronomy \& Astrophysics, University of Toronto, 50 St. George Street, Toronto, ON M5S 3H4, Canada}
\email{cissy.kuang@mail.utoronto.ca}

\email{alicia.savelli@utoronto.ca}

\begin{abstract}

How galaxies assemble depends on processes ranging from small-scale baryonic physics to large-scale cosmological assembly, which are difficult to disentangle with observations alone. Using $\sim$63,000 simulated galaxies from seven cosmological simulations spanning hydrodynamical, semi-analytic, and semi-empirical models, we explore how specific star formation histories (sSFHs) vary across modelling assumptions. Using non-linear dimensionality reduction (UMAP), we find that sSFH shape variation organizes along a few smooth axes in a 2D embedding, broadly corresponding to (1) peak timing, (2) peak sharpness, and (3) subsequent quenching rate. We identify sSFH shapes that many simulations share (``common''), well described to first order by dark matter halo assembly, and a few tied to individual simulations (``unique''), which likely reflect subgrid and parametrization choices such as merger-triggered burst prescriptions. A notable exception is a common population of rapidly quenched galaxies found across most simulations, likely driven by e.g., active galactic nucleus (AGN) feedback, that halo assembly alone cannot reproduce. The origin of the most distinctive unique shape, an Illustris-dominated history with an early peak and slow decline, remains unclear, but we test the effects of short-term variability. With a multi-objective auto-encoder neural network, we recover the simulation that a given sSFH is drawn from with $\sim$95\% out-of-sample accuracy; ablation tests indicate this information is embedded in long-timescale sSFH morphology rather than short-term variability. These results provide avenues for theoretical and observational follow-up connecting star formation histories to subgrid physical models.

\end{abstract}


\section{Introduction} \label{sec:intro}

\subsection{The Convergence of Cosmological Simulations}
Over the past decade, cosmological hydrodynamical simulations have evolved into essential laboratories for understanding galaxy formation \citep{Vogelsberger2020}. Modern simulation suites, including EAGLE \citep{Crain2015,Schaye2015}, IllustrisTNG \citep{Weinberger2017,Pillepich2018}, and SIMBA \citep{Dave2019}, have achieved remarkable success in reproducing the fundamental observables of the local Universe ($z=0$). These simulations employ vastly different numerical methods -- EAGLE uses Smoothed Particle Hydrodynamics (SPH), IllustrisTNG employs the moving-mesh code AREPO, and SIMBA utilizes GIZMO's meshless-finite-mass solver -- yet they broadly converge on key static properties, such as the Stellar Mass Function (SMF) and the existence of a tight Star Forming Main Sequence (SFMS), despite differences in its precise normalization \citep[e.g.,][]{Hahn2019, Iyer2020}.

This convergence, however, presents a profound theoretical challenge: the problem of \textit{equifinality}. Borrowed from systems theory and hydrology, equifinality refers to the phenomenon whereby multiple distinct pathways -- with different initial conditions and governing processes -- can lead to observationally indistinguishable endpoints \citep{vonBertalanffy1968, Beven2006}. In the context of galaxy formation, equifinality manifests as the inability to uniquely constrain subgrid physics from $z=0$ observables alone: the calibration process that tunes simulations to match the SMF and SFMS admits multiple solutions \citep{Crain2015}. Two galaxies may appear identical in their integrated stellar mass and star formation rate today, yet may have arrived at this state via radically different evolutionary pathways dictated by their specific feedback implementations.

The subgrid prescriptions used to model stellar and Active Galactic Nucleus (AGN) feedback illustrate this degeneracy. EAGLE relies on stochastic thermal energy injection to regulate star formation \citep{Crain2015,Schaye2015}, while SIMBA implements kinetic feedback with explicit jet modes that can reach velocities of $\sim 8000$ km s$^{-1}$ at low Eddington ratios \citep{Dave2019}, and IllustrisTNG utilizes a dual-mode thermal/kinetic AGN feedback scheme \citep{Weinberger2017}. Recent controlled experiments have made this equifinality explicit by resimulating the same halo under varied subgrid prescriptions. For a fixed implementation, varying the combination of star-formation and supernova feedback measurably alters the resulting dark matter halo structure even when the galaxies reach comparable stellar masses \citep{NunezCastineyra2023}; holding the rest of the physics fixed and switching only the supernova feedback coupling channel -- mechanical, hybrid thermal/mechanical, or delayed cooling -- produces Milky Way-mass galaxies with matched stellar masses yet fundamentally different circumgalactic media, metal distributions, and feedback modes \citep{Rey_2025}. Broader multi-code comparisons reach a similar conclusion at larger scales, with different hydrodynamics methods and subgrid implementations applied to a common initial condition producing qualitatively different disc morphologies in the Milky Way-mass progenitor over $1 < z < 5$ despite comparable integrated stellar masses \citep{Jung_2025}. Complementing these single-halo controlled experiments, the CAMELS project \citep{VillaescusaNavarro_2021, VillaescusaNavarro_2023, Ni2023}, which systematically varies both cosmological and astrophysical parameters across thousands of simulations, has revealed that machine learning models trained on one simulation suite fail to generalize to others \citep{Jo2025}. This ``domain generalization problem" highlights systematic differences between implementations that persist despite calibration to similar $z=0$ observables. To break these degeneracies and constrain the underlying physics, we must move beyond static scaling relations and examine the integrated dynamical record of galaxy growth: the Star Formation History (SFH).

\subsection{Star Formation Histories as Dynamical Fingerprints}
A galaxy's SFH is not merely a record of mass assembly; it is a time-resolved map of the competition between gravitational collapse and baryonic regulation. The variability and shape of SFHs encode specific physical processes acting on distinct timescales. By analyzing the Power Spectral Density (PSD) of star formation, \citet{Iyer2020} demonstrated that SFH variability is scale-dependent: on long timescales ($\gtrsim 2$~Gyr), SFHs are coherent with the dark matter accretion histories of their parent halos, governed largely by cosmology and gravity. On intermediate and short timescales ($\lesssim 1$~Gyr), however, SFHs decouple from the dark matter halo, driven instead by the complex interplay of gas cooling, baryon cycling, and feedback \citep{Tacchella_2020, Caplar2019}. The extended PSD framework of \citet{Iyer2024}, incorporating physically motivated Gaussian process kernels, further predicts that this high-frequency variability should be enhanced at early cosmic times.

Consequently, the full temporal ``shape" of an sSFH -- encompassing both its long-timescale morphology and any short-term burstiness -- may act as a fingerprint of the underlying subgrid physics \citep{Iyer2020}. Simulations with efficient gas ejection may produce SFHs with sharp truncations, while those with gentler feedback coupling may produce smoother histories \citep{Iyer2025b}. The impact of these physical choices is evident even when controlling for galaxy type: \citet{Savelli2025} found that when selecting for Milky Way analogues based on $z=0$ stellar mass and SFR, the evolutionary tracks taken to reach that state differ significantly between simulations.


JWST has revealed bursty and rapidly quenched star formation at $z > 4$ \citep{Looser2024,Endsley2025,Carnall2023}, placing SFHs at the centre of current tests of galaxy formation models. At lower redshift, non-parametric spectral energy distribution (SED) fitting now recovers the Gyr-scale shape of individual galaxies' SFHs \citep{Pacifici2016,Leja_2019,Carnall_2019}, so differences between simulations in long-timescale SFH shape are, in principle, observable. Characterizing which SFH shapes the simulations agree on, and which are specific to one model, is a prerequisite for using these observations to discriminate between models.

\subsection{A Manifold Learning Approach}
While the diagnostic power of SFHs is clear, comparing them across multiple simulation suites remains a methodological challenge. SFHs are high-dimensional time-series data, often comprising over a hundred time steps per galaxy. Traditional analyses have relied on summary statistics -- such as half-mass times ($t_{50}$), formation timescales ($\tau$), or quenching timescales -- to reduce this dimensionality. However, this compression can discard the detailed temporal structure of the sSFH that may carry distinguishing signatures of feedback between physical models \citep{Iyer2020, Pacifici2016}.

In this work, we present a comprehensive comparative analysis of $\sim 63{,}000$ galaxies across seven major simulation suites: EAGLE \citep{Crain2015, Schaye2015}, Illustris \citep{Vogelsberger2014}, IllustrisTNG \citep{Weinberger2017, Pillepich2018}, MUFASA \citep{Dave2016}, SIMBA \citep{Dave2019}, the Santa Cruz Semi-Analytic Model \citep[SC-SAM;][]{Somerville_etal_2008}, and the empirical UniverseMachine \citep{Behroozi_2019}. We additionally include three zoom simulations for reference -- FIRE-2 \citep{Hopkins_2018}, g14 \citep{Governato_etal_2012, Munshi_etal_2013, Brooks_Zolotov_2014}, and Marvel/Justice League \citep{Applebaum2021} -- though their smaller sample sizes preclude their use in our primary statistical analysis. Rather than reducing SFHs to scalars, we employ dimensionality reduction and deep learning techniques to analyze the full temporal structure of the galaxy histories. We utilize Uniform Manifold Approximation and Projection \citep[UMAP;][]{umap, umap2} to project the high-dimensional space of SFHs into a lower-dimensional landscape, allowing us to visualize the topology of galaxy formation models. This reveals regions of \textit{commonality}, where different physical models converge on the same evolutionary paths driven by gravity, and regions of \textit{uniqueness}, where specific subgrid recipes produce distinct classes of galaxies.

We further validate this visual mapping using an auto-encoder framework that compresses each sSFH into a low-dimensional latent representation from which the simulation of origin is classified. By comparing the classification accuracy of unsupervised embeddings (UMAP) against this learned latent space, we determine the extent to which subgrid physics leaves a unique, recoverable fingerprint on the history of a galaxy. This analysis provides a new framework for understanding where our current models of galaxy formation agree, where they diverge, and how future observations might distinguish between them.

\subsection{Goals and outline of this work}
The goal of this paper is to determine which SFH shapes are ``common" across several cosmological simulations and which are ``unique" to individual simulations or simulation families, and subsequently to investigate what underlying subgrid processes may be contributing to sculpting these shapes.  We build on the SFH catalogue of \citet{Iyer2020}, who introduced the power spectral density (PSD) of star formation as a framework for quantifying how strongly SFHs vary on different timescales, showing that models differ widely in this variability. Although we use the same simulations and SFHs, our question is complementary.  Rather than studying the SFH variability, in this work we retain the full temporal shape of every sSFH (the timing and sharpness of its peak, and the rate of its subsequent quenching), and ask where in this shape space the simulations converge and where they diverge.

The remainder of this paper is organized as follows. In Section~\ref{sec:data}, we briefly describe the relevant subgrid physics in each of the seven cosmological simulations and the preparation of their sSFHs. In Section~\ref{sec:methods}, we introduce the UMAP dimensionality reduction of the sSFHs and explore the resulting projection space. In Section~\ref{sec:results}, we quantify the diversity of SFHs across simulations using Simulation-of-Origin Probabilities and the Shannon entropy, which together separate common from unique shapes. In Section~\ref{sec:discussion}, we interpret these regions in terms of halo assembly and baryonic subgrid physics, test the information content of the sSFHs with a multi-objective auto-encoder, apply our framework in the context of Milky Way analogue evolution, and discuss caveats. We summarize our conclusions in Section~\ref{sec:conclusions}.

\section{Data} \label{sec:data}
Cosmological simulations come in a variety of ``flavours", each employing their own physical and numerical models on a range of scales.  Two popular and traditional approaches are \textit{\mbox{(magneto-)}hydrodynamical models}, which simultaneously solve for gravity using $N$-body methods and evolve the gas coupled to the gravitational field (and electric and magnetic fields) as a fluid, and \textit{semi-analytical models}, which evolve the dark matter and gas components separately using simplified flow equations describing bulk motions in the simulation box.  A more recent addition to the group is \textit{semi-empirical models}, which are tuned specifically to match a range of observations.   

Our dataset comes from \citet{Iyer2020} and is comprised of galaxies from seven cosmological simulations with various physical and numerical models in our main analysis, including \mbox{(magneto-)}hydrodynamic simulations (EAGLE, Illustris, IllustrisTNG, Mufasa, and Simba), semi-analytical models (Santa-Cruz SAM), and semi-empirical models (Universe Machine).  We consider only the fiducial run from each simulation since some simulations also provide runs with variations on the parameters of the subgrid physical models.  The specifications and properties of each simulation are summarized in Table~\ref{tab:sim_specs}.  We additionally include a handful of galaxies from \textit{zoom simulations}, which evolve galaxies with higher resolution and thus more explicit subgrid prescriptions, in an extended analysis in Appendix~\ref{app:zoom}.  

In this section, we briefly outline the simulation properties and relevant subgrid models in Section~\ref{sec:sims}, and summarize the methods used to prepare the original data from \citet{Iyer2020} for our own analysis in Section~\ref{sec:manipulation}.

\subsection{Simulation physics} \label{sec:sims}


We examine a range of fully \mbox{(magneto-)}hydrodynamic, semi-analytic, and semi-empirical models in this study. In this section, we introduce the models that this paper explores and cite the specific physical prescriptions they implement.

\begin{table*}[]
    \centering
    \begin{tabular}{c|cc|ccc|c}
    \hline
         Type & Simulation Name & Run Name & Box Length & $m_\textrm{dm}$ & $m_\textrm{sp}$ & $n_\textrm{gal}$ \\ 
         &&& [Mpc] & $\left[10^6\MSun\right]$ & $\left[10^6\MSun\right]$ \\
         \hline\hline
         Hydro & EAGLE & Ref-L100N1504 & 100 & 9.7 & 1.81 & 7445 \\ 
         Hydro &  Illustris & Illustris-1  & 106.5 & 6.26 & 1.26 & 19354 \\
         Hydro & IllustrisTNG & TNG100-1 & 110.7 & 7.5 & 1.4 & 12220 \\ 
         Hydro & Mufasa & m50n512 & 50 & 96 & 48 & 1900 \\
         Hydro & Simba & m100n1024 & 100 & 96 & 18 & 1982 \\
         \hline
         SAM & Santa-Cruz SAM & - & 100 & 203.7 & N/A & 12821 \\
         \hline
         SEM & UniverseMachine & - & 70.3 & 203.7 & N/A & 7361 \\
         \hline
    \end{tabular}
    \caption{Brief descriptions of the models studied in this work. For further details, please see \citet{Iyer2020}. }
    \label{tab:sim_specs}
\end{table*}

\begin{enumerate}
    \item\textbf{EAGLE}: (\cite{Schaye2015}, \cite{Crain2015}, \cite{McAlpine_etal_2015}, \cite{Shaller_etal_2015}, \cite{Wiersma_Shaye_Smith_2009}, \cite{Wiersma_etal_2009}, \cite{Durier_DallaVecchia_2012})
    EAGLE (Evolution and Assembly of GaLaxies and their Environments) is a numerical hydrodynamic simulation. Gravity, hydrodynamics, and thermodynamics are solved using a modified version of GADGET-3, (\cite{Springel_2005}), which is an N-body Tree-Particle-Mesh smoothed particle hydrodynamics (SPH) code. Star formation physics is modelled to follow \cite{Shaye_DallaVecchia_2008} and \cite{Shaye_2004} using a metallicity-dependent density threshold to determine where gas particles can be stochastically converted into star particles following the KS relation, and isotropic stochastic thermal stellar feedback from \cite{DallaVecchia_Shaye_2012} that can drive large-scale galactic outflows at high SFR. AGN feedback is implemented through a single-mode thermal feedback channel, with the fraction of energy that couples to the ISM calibrated to reproduce the $M_*-M_{BH}$ relation at $z\sim 0$, mimicking radio- and quasar-like behaviour depending on the SMBH accretion rate \cite{Springel_etal_2005a}, \cite{Booth_Shaye_2009}, and \cite{RosasGuevara_etal_2015}.
    
    \item\textbf{Mufasa}: (\cite{Dave2016}, \cite{Hopkins_2015})
    Mufasa is a numerical hydrodynamic simulation that uses the Meshless Finite Mass (MFM) solver GIZMO (\cite{Hopkins_2014}) code to model cosmological gravity and hydrodynamics. It also includes prescriptions from Grackle (\cite{Smith_etal_2017}) for cooling and heating.
    Star formation is implemented stochastically from gas particles using a $H_2$ density-based threshold \cite{Krumholz_McKee_Tumlinson_2009}, \cite{Krumholz_Gnedin_2011}, \cite{Thompson_etal_2014}.
    Stellar feedback drives two-phase winds (combining type-II supernovae, radiation pressure, and stellar winds), with mass loading factors and wind speeds adopted from the FIRE simulations (\cite{Hopkins_etal_2014}, \cite{Muratov_etal_2015}).  Stellar feedback follows \cite{Springel_Hernquist_2003} and \cite{Oppenheimer_Dave_2006}. While AGN feedback is not explicitly modeled, the solution implemented for quenching massive galaxies invokes `maintenance-mode' feedback by keeping all the gas (except self-shielded gas) in halos heated \cite{gabor_Dave_2012} and \cite{Gabor_Dave_2015}.
    
    \item\textbf{Simba}: (\cite{Dave2019}, \cite{Hopkins_2014}, \cite{Hopkins_2017})
    Simba is the successor of the Mufasa simulation. Simba follows similar methods to Mufasa, implementing MFM cosmological gravity and hydrodynamics code GIZMO. Star formation follows the same stochastic $H_2$-based model as Mufasa, with the SFR given by the ratio of the $H_2$ density to the local dynamical timescale.
    The major goal of the Simba simulation is to address the lack of AGN feedback prescriptions in the Mufasa simulation, in addition to other changes, like new two-phase winds using the FIRE simulation's updated mass loading factor scalings (\cite{Alcazar_etal_2017a}), similar to IllustrisTNG but with slightly lower wind velocities. Wind particles are briefly decoupled from hydrodynamics and radiative cooling.
    AGN accretion is modeled using a torque-limited model \cite{Alcazar_etal_2017b}, with the resulting feedback including thermal and kinetic modes with variable outflow velocities. Feedback operates via a jet mode at low accretion rates (responsible for a majority of quenching) and a radiative/X-ray mode at higher accretion rates. \cite{Dave2019, RodriguezMontero_etal_2019}.
    
    \item\textbf{Illustris}: (\cite{Vogelsberger2014}, \cite{Vogelsberger_etal_2014b}, \cite{Genel_etal_2014}, \cite{Sijacki_etal_2015})
    Illustris is a numerical hydrodynamic simulation. It models galaxy formation using the moving mesh code AREPO (\cite{Springel_2011}). This simulation provides physical process prescriptions for primordial cooling, metal-line cooling, stellar evolution, stellar feedback, gas recycling, chemical enrichment, BH growth, and AGN feedback. Star formation is implemented stochastically \cite{Vogelsberger_etal_2013} and \cite{Torrey_etal_2014} above a critical hydrogen number density, using a sub-grid recipe with an effective equation of state \cite{Springel_Hernquist_2003} that reproduces the  observed Kennicutt-Schmidt relation (\cite{Schmidt_1959}, \cite{Kennicutt_1989}). Stellar feedback prescriptions from \cite{Vogelsberger_etal_2013} drive kinetic winds via hydrodynamically decoupled wind particles that recouple to the gas once they leave the dense ISM. The BH and AGN feedback prescriptions follow from \cite{Springel_etal_2005b}, \cite{Sijacki_etal_2007}, \cite{DiMatteo_etal_2005}, and \cite{DiMatteo_etal_2008}, operating in two modes: a thermal (quasar) mode and a kinetic (radio) mode, with quenching driven primarily by radio-mode feedback.

    \item\textbf{IllustrisTNG}: (\cite{Pillepich_etal_2018b}, \cite{Weinberger2017}, \cite{Marinacci_etal_2018}, \cite{Naiman_etal_2018}, \cite{Nelson_etal_2018a}, \cite{Springel_etal_2018})
    IllustrisTNG is The Next Generation (TNG) of the Illustris simulation. It is a numerical hydrodynamic simulation. While it implements many prescriptions used in the Illustris simulation, TNG includes many new features, like injecting winds isotropically, and a new form of BH-driven kinetic feedback (\cite{Weinberger2017}, \cite{Weinberger_etal_2018}). While star formation is implemented identically to Illustris, the stellar feedback is significantly updated, with the wind speed scaling with the local 1D dark matter velocity dispersion, and a redshift-dependent normalization tied to halo mass growth \citep{Pillepich2018}. AGN feedback is still two-mode, with a pure thermal (quasar) mode at high accretion rates and a kinetic wind mode at low accretion rates (radio mode) \citep{Weinberger2017}. The latter is primarily responsible for quenching galaxies at high stellar masses.
    
    \item\textbf{Santa-Cruz SAM (SC-SAM)}: (\cite{Somerville_etal_2008}, \cite{Somerville_etal_2015}, \cite{Porter_etal_2014}, \cite{Brennan_etal_2017})
    Santa-Cruz SAM is a semi-analytical model. Santa-Cruz SAM makes use of the Bolshoi-Planck dark matter (DM) merger-tree simulations (\cite{Klypin_etal_2011}, \cite{Klypin_etal_2016}, \cite{RodriguezPuebla2016a}) to construct its galaxies. Santa-Cruz SAM implements physical process prescriptions for the hierarchical growth of structure, gas heating, gas cooling, star formation, stellar evolution, supernova feedback, starbursts, and morphological transformations due to mergers and disc instabilities. Star formation follows a disc mode following the KS relation, and a starburst mode triggered by mergers or disc instabilities. Stellar feedback drives winds that eject cold gas from the galaxy reservoir, with heated gas either retained in the halo potential well or ejected into the IGM. A multiphase ISM model tracks the cold gas availability for star formation. AGN feedback (driven by mergers or disc instabilities) provides the dominant quenching mechanism for massive galaxies, along with virial shock heating, following prescriptions for SMBH growth and feedback detailed in \citet{Somerville_etal_2008, Brennan_etal_2017}.
    
    \item\textbf{UniverseMachine (UM)}: (\cite{Behroozi_etal_2013}, \cite{Behroozi_2019})
    UniverseMachine is a semi-empirical model. This simulation uses the Bolshoi-Planck DM simulations (\cite{Klypin_etal_2011}, \cite{Klypin_etal_2016}, \cite{RodriguezPuebla2016a}) to obtain DM halo properties and assembly histories. This simulation was run using the ART code (\cite{Kravtsov_etal_1997}, \cite{Kravtsov_Klypin_1999}). Halo-finding and merger tree construction was done using the \texttt{ROCKSTAR} (\cite{Behroozi_etal_2013b}) and \texttt{CONSISTENT-TREES} (\cite{Behroozi_etal_2013d}) codes. UniverseMachine assigns SFRs to galaxies as a function of host halo potential well depth, assembly history and redshift. Stellar and AGN feedback are both not explicitly modelled, and are instead inferred from the SFR-halo property correlations constrained using MCMC optimization against a wide range of observations across $0<z<10$. The resulting quenching timescales and fractions, and their environmental and mass dependence, provide empirical constraints on the effects of feedback.
\end{enumerate}

\subsection{Data manipulation}\label{sec:manipulation}
The SFHs presented in this work are those extracted from the $z\sim 0$ simulation snapshots by \citet{Iyer2020}, in which the authors compute the SFH for a given subhalo through a mass-weighted binning of the associated star particles in bins of $\Delta t=100$~Myr for the \mbox{(magneto-)}hydrodynamic simulations, accounting for mass-loss to estimate the total stellar mass formed. For SC-SAM and UM, the SFHs per galaxy are interpolated to match the same time grid.  While this summary of their methodology is brief, interested readers are invited to consult the original text for a more detailed description, specifically their Section 2.2.  In this work, we then employ a series of cuts, interpolations, and normalizations on the extracted SFHs in order to prepare them for our own analysis, described as follows.  

As in \citet{Iyer2020}, we apply a cut on the data to include only galaxies above a present-day stellar mass threshold of $10^9\,\MSun$ in order to partially mitigate resolution effects.  In the case of Mufasa and Simba which have lower resolution (see Table~\ref{tab:sim_specs}), this threshold is raised to $10^{10}\,\MSun$.  
In this work, we also exclude any galaxies that have $\SFR=0$ at every snapshot, as these likely result from a glitch and will not benefit our analyses.  
The final number of galaxies from each simulation included in our dataset after these cuts is shown in the final column of Table~\ref{tab:sim_specs}, for a total sample size of 63,083 galaxies. 

\begin{figure*}[htbp!]
    \centering
    \includegraphics[width=\textwidth]{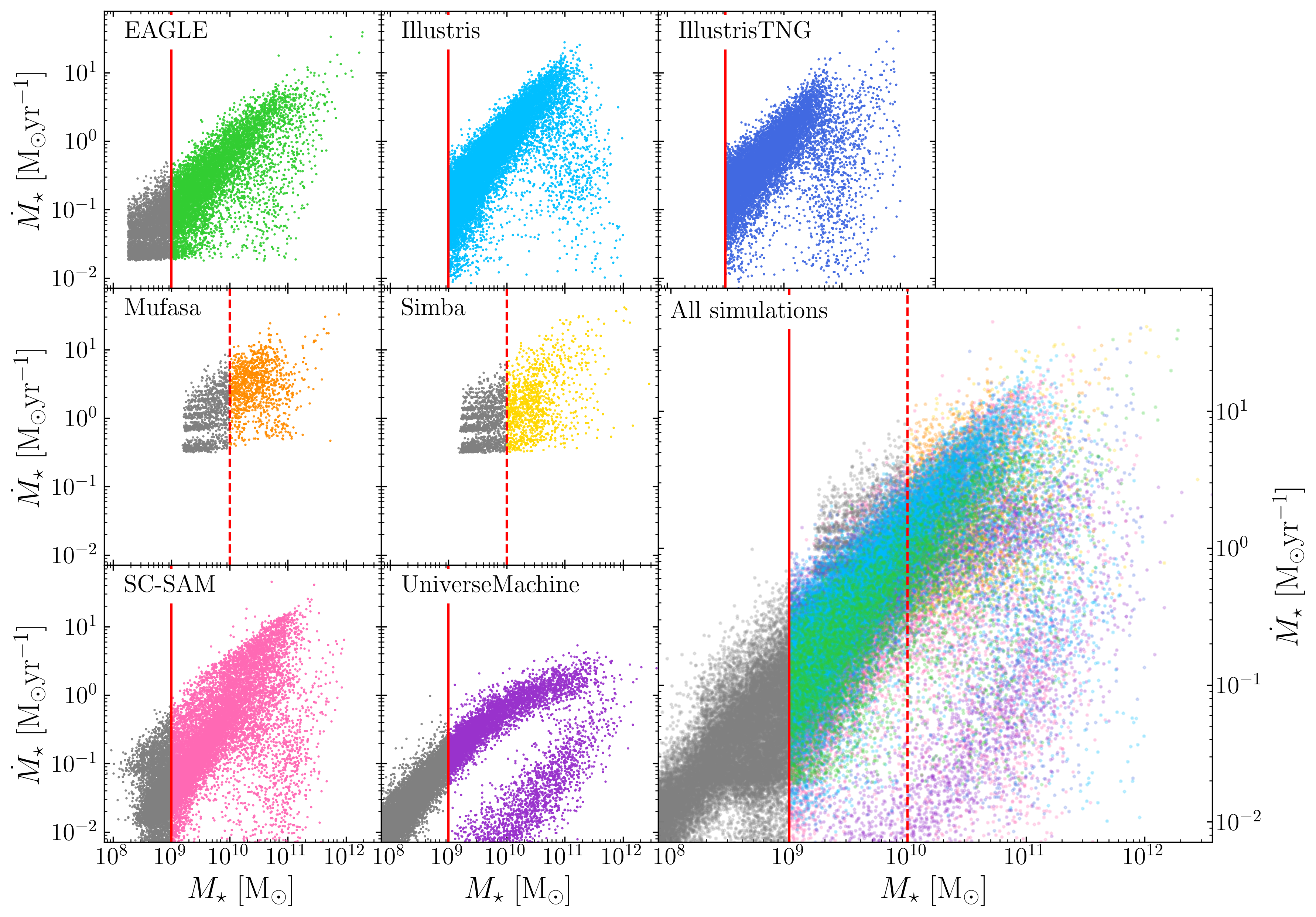}
    \caption{The $z=0$ star-forming main sequence (SFMS) for each of the seven simulations in our sample (EAGLE: green, Illustris: light blue, IllustrisTNG: dark blue, Mufasa: orange, Simba: yellow, SC-SAM: pink, UniverseMachine: violet), with the combined distribution shown in the large \textbf{bottom-right} panel. Coloured points lie above each simulation's mass cut ($\SM > 10^9\,\MSun$, marked with solid red; $\SM > 10^{10}\,\MSun$ for Mufasa and Simba, marked with dashed red) and are included in our final sample; grey points lie below the cut and are excluded. The per-simulation colour scheme shown here is reused throughout this work; see Section~\ref{sec:manipulation} for details on the cuts, per-simulation sample sizes, and visualization conventions. The seven simulations produce roughly consistent galaxy populations at $z=0$ on the scaling observables against which they were calibrated -- though features such as the secondary-sequence structure visible in UniverseMachine already hint at implementation-level differences. This makes it unclear if the evolutionary pathways leading to these consistent present-day populations are themselves consistent, or whether distinct histories are hidden behind their matching $z=0$ properties.
    }
    \label{fig:SFMS}
\end{figure*}

To visualize this final dataset, we show the $z=0$ star-forming main sequence (SFMS; the relationship between a galaxy's stellar mass $\SM$ and SFR $\SFR$) for each simulation, individually and combined, in Figure~\ref{fig:SFMS}.  The vertical red lines represent the mass cuts, with only coloured points to the right of the lines being included in our final dataset.  The horizontal striping features in the grey points exhibited particularly by Mufasa, Simba, and EAGLE are artifacts of the resolution limits, and thus demonstrate the need for the application of such mass thresholds.  The colours assigned to each simulation in the figure (EAGLE: green, Illustris: light blue, IllustrisTNG: dark blue, Mufasa: orange, Simba: yellow, SC-SAM: pink, and UM: violet) are used throughout this work to match galaxies to their home simulation.   The apparent absence of grey points below the mass cut for Illustris and IllustrisTNG reflects our initial database query, which did not return galaxies below $10^9\,\MSun$ from these two simulations -- galaxies at that mass exist but have too few star particles to resolve their SFHs. For SC-SAM and UM, a lower bound on the $\SFR$ axis has been applied in Figure~\ref{fig:SFMS} for visualization purposes only; the full sample extends to lower SFR values.

Our methodology (as described in Section~\ref{sec:methods}) requires a comparison of the specific SFR (sSFR) of each galaxy on a snapshot-by-snapshot basis with equal time intervals between snapshots over an identical time range.  The extracted SFHs from \citet{Iyer2020} do not share a perfectly identical time grid, as the simulations are adopt slightly different cosmologies and thus have different ages at z=0 (e.g., EAGLE is evolved to 13.6~Gyr, while UM is evolved to 13.8~Gyr).  We first compute the specific SFHs (sSFHs) by normalizing via the time-integrated SFHs, as evaluated numerically over each simulation's unique time grid.  We denote the time-integrated SFHs as $m_\star$, rather than $M_\star$ as used to denote stellar mass, as these two calculations need not yield the same value due to effects such as mass loss.  We then standardize the time grids by interpolating the sSFHs over a range of 0 to 13.8 Gyr in intervals of 100 Myr, and perform a time cut to only include snapshots within the bounds of the original time grids, such that $0.1~\mathrm{Gyr}\leq t\leq 13.6~\mathrm{Gyr}$.  

\begin{figure*}[!ht]
    \centering
    \includegraphics[width=0.85\textwidth]{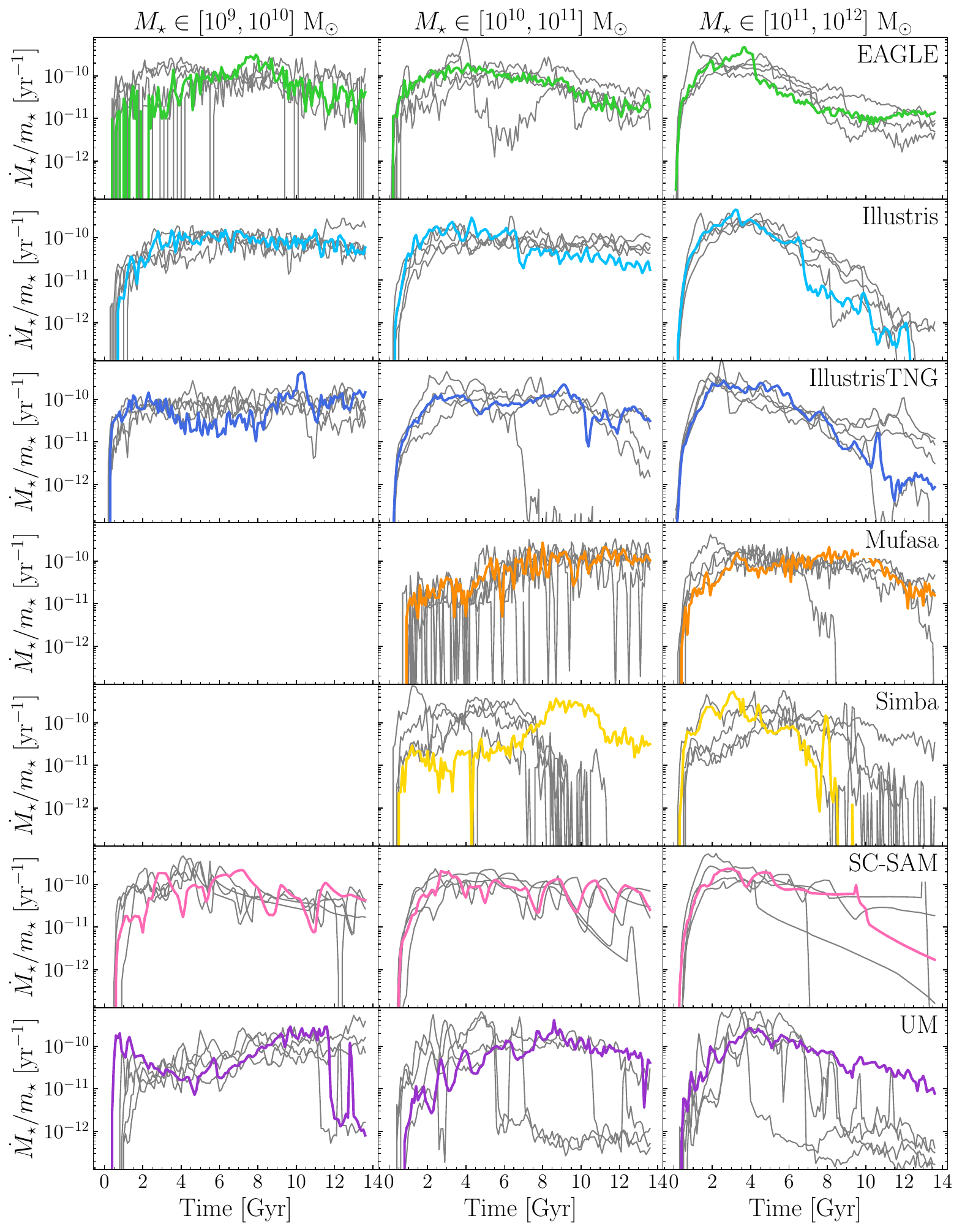}
    \caption{Sample specific star formation histories (sSFHs) drawn from each simulation, after the data preparation described in Section~\ref{sec:manipulation}, in 1 dex stellar mass bins $\in[10^9,10^{12}]\,\MSun$ increasing from left to right; the leftmost bins for Mufasa and Simba are left blank by construction.  From these samples it is already clear that similar sSFHs profiles can be produced across all simulations, while some simulations are able to produce more distinct profiles, such as the burstiness on short time scales exhibited by Mufasa, Simba, and EAGLE \citep{Iyer2020}.  The mass binning also demonstrates why grouping by similar $\SM$ alone is not sufficient to adequately compare sSFH profiles -- as especially evident in the middle panels for Universe Machine and IllustrisTNG, showing sSFHs with substantially different growth rates, peak times, and quenching rates -- thus motivating the need for a more sophisticated method of sorting the data and thus the shape-based grouping introduced in Section~\ref{sec:methods}.    }
    \label{fig:sSFH}
    \end{figure*}

Figure~\ref{fig:sSFH} shows a selection of the resulting sSFHs sampled from 1 dex mass bins in the range $10^9\,\MSun$ to $10^{12}\,\MSun$.  One may naively suggest that the galaxies be sorted by simulation or binned by stellar mass for means of comparing their sSFHs; however, the pictured sSFH profiles already show a diverse group of shapes (i.e., growth rates, peak times, quenching rates) and textures (i.e., burstiness), even within the same mass bin for a given simulation.  Thus, the requirement to compare the sSFHs of each galaxy over a common time grid is well-motivated. 

\section{Methodology for comparing star formation histories\label{sec:methods}}
In this section, we develop our methodology and means for comparison of our sSFH sample.  In Section~\ref{sec:umap} we apply a dimensionality reduction technique, and explore the resulting projection space in Section~\ref{sec:projection}.  As a complementary quantitative probe of the same manifold, we additionally train a multi-objective auto-encoder on our sSFH sample; the preprocessing, architecture, and training details are deferred to Appendix~\ref{app:zehao}.

\subsection{Dimensionality reduction\label{sec:umap}}
As stated in Section~\ref{sec:manipulation}, we wish to compare the sSFHs on a snapshot-by-snapshot basis.  Since every sSFH contains 136 snapshots, we can treat each one as a 136-dimensional object.  Dimensionality reduction techniques such as UMAP~\citep{McInnes2020umap}, t-SNE \citep{vanderMaaten2008}, and PCA \citep{Pearson1901, Jolliffe2002} have recently become popular means for qualitative analysis in studies involving high-dimensional datasets \citep{Iyer2020, Reis2018, Portillo2020}, as they use machine learning methods to reduce the data to a 2- or 3D projection, preserving the remaining information in a latent space.  These low-dimensionality projections allow for simplified data visualization and exploration while maintaining information about the underlying structure.  We thus apply such a technique to our 136D sSFHs, sorting them into a 2D projection that we use to qualitatively compare similarities and differences between each sSFH profile on the basis of various metrics.  

\begin{figure*}[ht!]
    \centering
    \includegraphics[width=\textwidth]{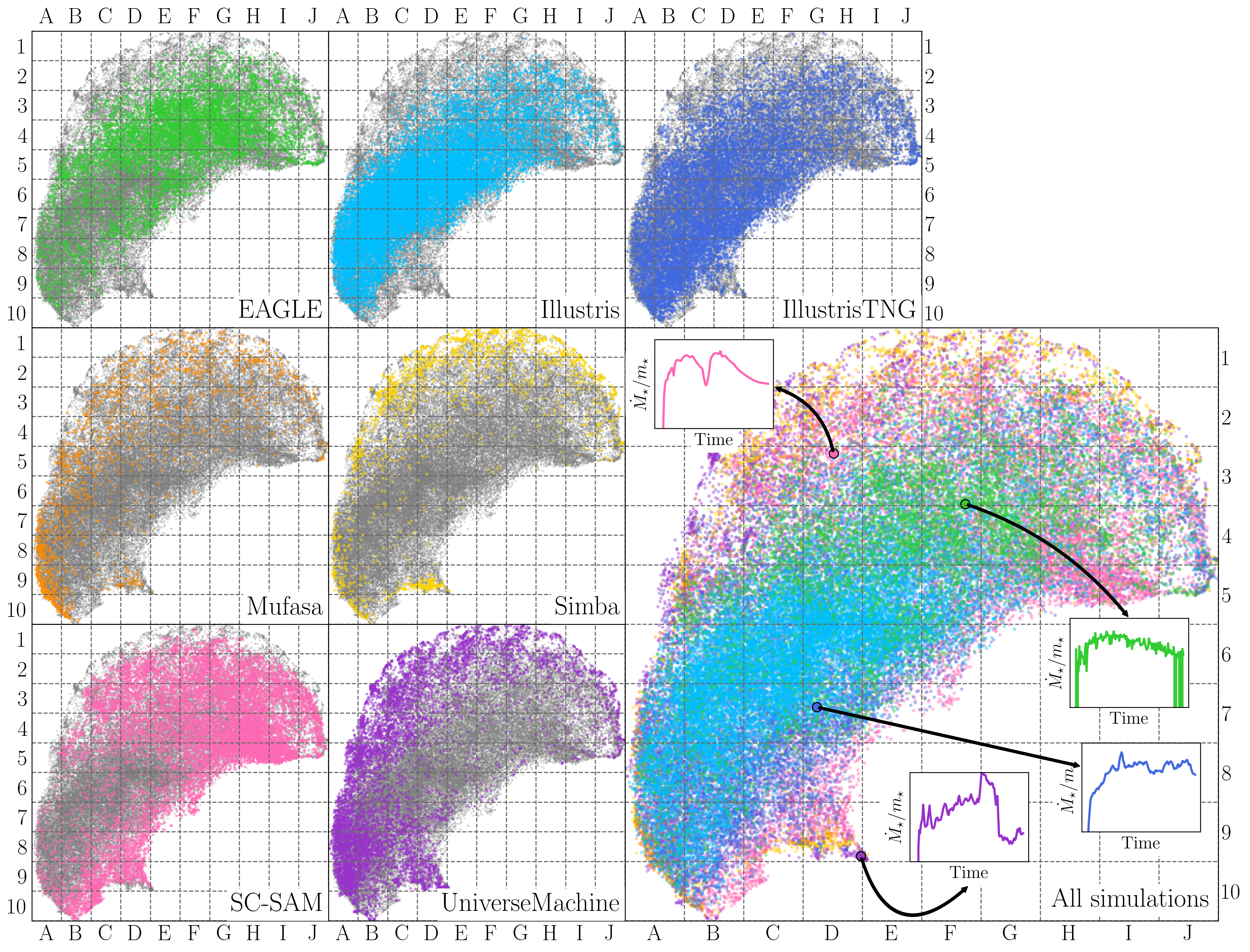}
    \caption{
    UMAP projection of the sample of sSFHs reduced to two dimensions; the general shape of which we refer to as a ``taco".  The layout and colours are the same as in Figure~\ref{fig:SFMS}, though here the coloured points in the individual simulation panels indicate the galaxies from the respective simulation, overlaying the galaxies from the remaining sample in grey.  Each point in the projection space represents one galaxy's sSFH, as illustrated by the insets corresponding to highlighted points in the combined-simulations bottom-right panel.  UMAP groups points in this space by how ``similar" a given sSFH is to its neighbours based on the 136-dimensional sSFHs (corresponding to 136 simulation time steps) supplied to the algorithm.  Thus, the horizontal and vertical axes are physically meaningless, and instead what is most informative is the clustering and spacing between points; two points close together represent galaxies with very similar sSFH profiles, with more distant points corresponding to distinct sSFH profiles.  As a means to compare clusters of sSFHs, we have grouped the galaxies into 100 bins in this projection space in the style of a ``battleship grid", identifying individual bins by an alpha-numeric pair (e.g., D7).  While there is some overlap between all simulations, there are certain regions of the taco dominated by certain (groups of) simulations, providing a hint into which sSFH profiles may be common amongst all simulations, and which may be unique to a given (group of) simulation(s).  The goal of this paper is to describe common versus unique sSFH structure across all simulations.
    }
    \label{fig:UMAP}
\end{figure*}

Our dimensionality reduction package of choice is UMAP \citep{McInnes2020umap}, as it offers the flexibility of fitting new data into an already existing projection space -- a feature we use to include the zoom simulations in Appendix~\ref{app:zoom}.  After feeding the normalized and interpolated sSFHs into UMAP, we retrieve the embedding shown in Figure~\ref{fig:UMAP}, where each point represents a sSFH in the projection space coloured by simulation.  This can be seen in the large, bottom right panel showing the UMAP of the sSFHs from all simulations combined, with insets tying highlighted points in the projection space to their corresponding sSFH.  

It is important to note that the $x$- and $y$-axes and the associated spatial positions of the points within the UMAP projection space do not necessarily hold physical meaning; rather, the relative spacing between points depicts the similarity of the corresponding sSFHs.  In other words, two points that are spatially close together on the UMAP projection represent two galaxies with similar structure in their sSFHs, and groupings of points represent groupings of sSFHs that all have similar structure.  Likewise, points or groups spread out on the UMAP indicate galaxies or groupings of galaxies with very different structure in their sSFHs.  UMAP instead determines implicit axes in the projection space to organize points based on likeness to their neighbours, which we discuss in more detail in Section~\ref{sec:projection}.  

In order to discuss structure and trends in the UMAP space without the support of any obvious axes system, we have separated the data into 100 equal-sized bins, as illustrated by the ``battleship gridding" system in Figure~\ref{fig:UMAP}, and we will refer to individual cells with an alpha-numeric pair (e.g., D7).  We use this battleship system for the remainder of this work not just to reference specific areas in the projection space, but also to help visualize trends in the data based on the cluster of points in a given cell.  To speak of regions in the projection space more broadly, we refer to its general shape as a ``taco". 

The smaller panels in Figure~\ref{fig:UMAP} show where the sSFHs from any given simulation lie overtop of the entire UMAP projection in grey.  The region of the taco occupied by some simulations seems constrained to a given area, while for other simulations, the associated sSFHs are more-or-less scattered across most of the taco.  Unsurprisingly, simulation families occupy much of the same regions of the taco, with Mufasa and Simba dominating the upper taco shell, while Illustris and TNG reside in the central and lower regions of the taco body.  EAGLE also has significant overlap with Illustris and TNG, presumably due to this trio of simulations having similar subgrid physics and simulation properties such as box-size and resolution.  

Interestingly, there is also considerable overlap between SC-SAM and EAGLE, as well as UM with Mufasa and Simba.  The ties between these hydrodynamic simulations with the SAM and SEM is less obvious, and will have to be explored in more detail later in this work.  Perhaps the most intriguing region of the projections space is the spike at the base of the taco, centred around cell D9. This protrusion signifies that the UMAP algorithm found notable differences between these sSFHs and the surrounding regions, which we will return to in our analysis in Section~\ref{sec:discussion}.

\subsection{Exploring the UMAP projection space\label{sec:projection}}
While the UMAP projection in Figure~\ref{fig:UMAP} offers a simple visualization of the distribution of sSFHs, further exploration of the projection space is needed to understand the layout of the distribution in terms of what trends the UMAP algorithm has incorporated into its embedding.  In other words, we wish to explain \textit{why} UMAP has created this particular embedding -- what characteristics of the sSFHs dictated their positioning in the projection space?  To answer this question, we use our battleship-style binning system to study trends in the shapes and related galaxy properties of the sSFHs in the projection space.

\begin{figure*}[hbtp!]
    \centering
    \includegraphics[width=\textwidth]{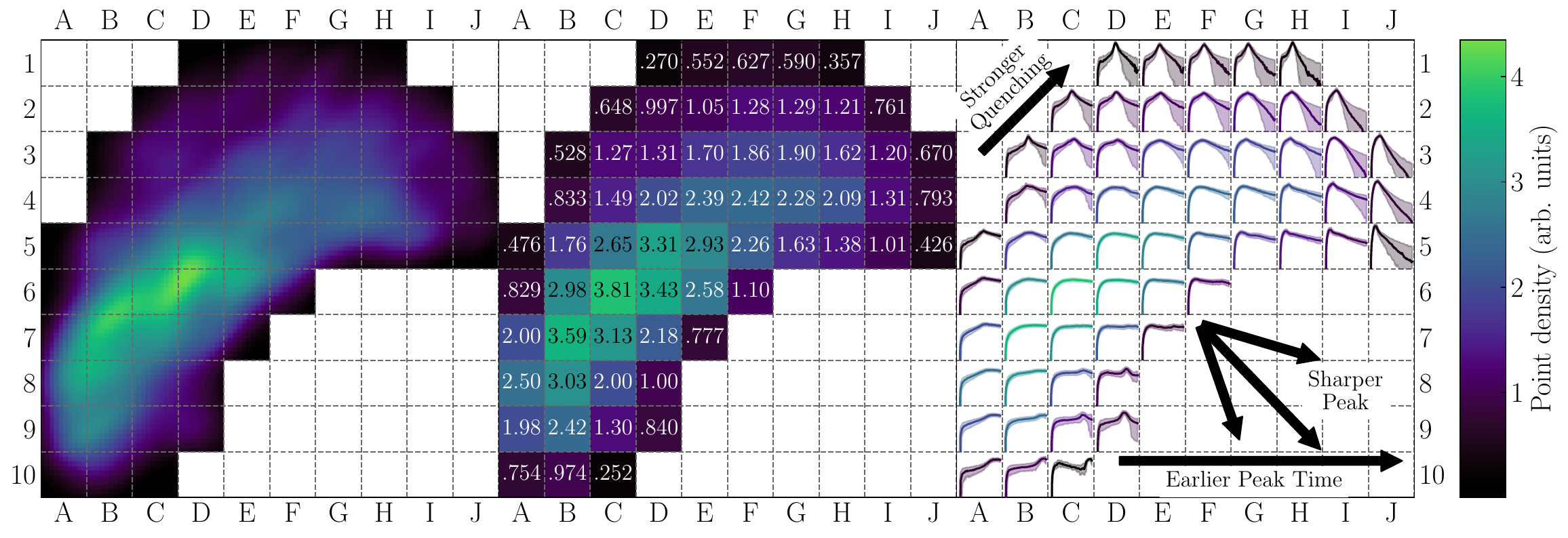}
    \caption{Density of points in the UMAP projection space, computed from a normalized Gaussian KDE applied over the projection space, including only grid cells with at least 100 data points. The \textbf{left} panel shows the density evaluated on a $100\times100$ mesh grid. The \textbf{centre} panel shows the same density averaged within each battleship-grid cell (value printed). The \textbf{right} panel shows each cell's median sSFH (solid line) with its 16--84 percentile range (shading), coloured by the cell-averaged density. This three-panel layout -- mesh or point-wise field, cell-averaged field, cell-median sSFH -- is reused throughout the paper as a template for any quantity mapped onto the UMAP projection. The point distribution peaks in the centre of the taco and decreases radially. The per-cell sSFHs reveal three implicit projection axes (arrows): earlier peak time from left to right, stronger quenching along the A10 $\to$ J1 diagonal, and a steeper peak moving radially outward from the central high-density region.
    }
    \label{fig:kde}
\end{figure*}
We first consider the density distribution of points in the UMAP projection space by calculating a Gaussian kernel density estimation (KDE) on the entire galaxy sample evaluated over a $100\times100$ mesh grid, normalized to integrate to $10^4$.  The resulting continuous point density distribution\footnote{Note that UMAP does not exactly preserve the density or distance of the original high-dimensional data, rather it preserves the general global structure \citep{umap2}.} is shown in Figure~\ref{fig:kde}.  The layout of Figure~\ref{fig:kde} includes 3 panels, the left displaying the density distribution over the $100\times100$ mesh grid, the middle marking the average point density per grid cell, and the right showing the average sSFH per grid cell coloured by the point density, with the solid line representing the median and the shaded region the 16$^\mathrm{th}$ -- 84$^\mathrm{th}$ percentiles.  Only cells containing more than 100 points are included in the figure to ensure a sufficient sample size for analysis.  This layout is repeated in all remaining figures in this work to illustrate how a given quantity changes over the UMAP projection space, and how it relates to the sSFH shape.

From the left and centre panels, the region of highest point density is the centre of the taco, with cell C6 having the most points, and density decreasing radially outwards to thin at the edges of the taco.  From the right panel, we can begin to decipher the trends in sSFH shape across the projection space.  We identify three clear implicit axes in the projection space: (1) the time of the peak in star formation occurs earlier when crossing directly from left to right, (2) the galaxies exhibit stronger quenching along the diagonal from A10 through J1, and (3) travelling radially outward from the centre of the taco the peak in star formation steepens.  These trends offer some insight into how the UMAP algorithm sorted the sSFHs in the projection space based on similarities in their shapes, though these axes are only meant to illustrate a qualitative analysis, and by no means do they exhibit all the intricacies of the UMAP algorithm.  

\begin{figure*}[hbtp!]
    \centering
    \includegraphics[width=\textwidth]{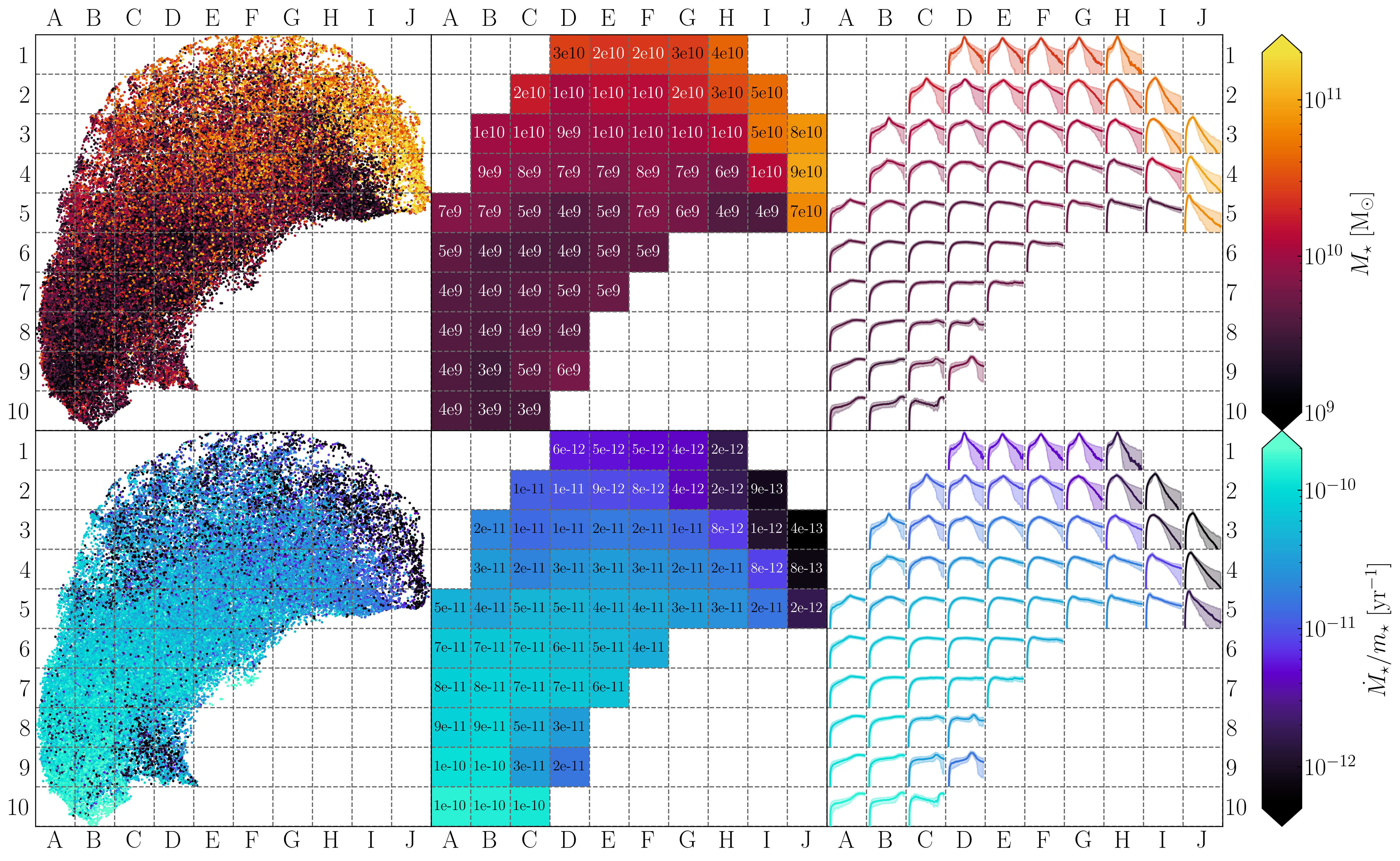}
    \caption{The $z=0$ stellar mass ($\SM$, \textbf{top row}) and specific star formation rate ($\sSFR$, \textbf{bottom row}) of our galaxy sample, shown in the same three-panel layout as Figure~\ref{fig:kde} (individual points, cell log-average with values printed, and cell-median sSFH coloured by the cell-average). Both quantities vary smoothly along the A10 $\to$ J1 diagonal of the taco, with $\SM$ increasing and $\sSFR$ decreasing. Combining these trends with the implicit axes of Figure~\ref{fig:kde} shows the upper-right end of this diagonal corresponds to galaxies that assembled rapidly and quenched sharply, while the lower-left end corresponds to slow-assembling, still-star-forming galaxies.  Ultimately, these trends are not surprising as present-day $\sSFR$ is the final value of each sSFH,  and SFH shape is known to correlate with stellar mass \citep{Pacifici2016, Iyer2020}.}
    \label{fig:sm/sfr}
\end{figure*}

We next search for trends in the $z=0$ galaxy properties.  Figure~\ref{fig:sm/sfr} shows the $z=0$ stellar mass ($\SM$, top row) and sSFR ($\sSFR$, bottom row) of each galaxy in the projection space, following the same layout as Figure~\ref{fig:kde}, with the left panels now showing the stellar mass and sSFR associated with each individual point, and the average values in the centre and right panels computed in log space.  Clear trends exist in both cases, with low $\SM$ (high $\sSFR$) in cell A10 -- showing late-peak, slow-quenching sSFHs -- and increasing (decreasing) along the diagonal towards J1 -- showing early-peak, rapid quenching sSFHs.

It is unsurprising for the UMAP algorithm to pick up on the $z=0$ sSFR, which is given by the last point in the sSFH. The trend in stellar mass across the projection space is slightly more surprising, since the sSFHs have been normalized by the time-integrated sSFH to eliminate mass-dependence, as this essentially gives the stellar mass; however, the sSFH shapes in the upper right (cells H1, I2, I3, J3, J4, J5) are typical of higher-mass galaxies, and the shapes in the lower left (cells A8, A9, A10, B8, B9, B10) are indicative of lower-mass galaxies \citep{Pacifici2016, Iyer2020}.

\section{Star formation history diversity across simulations \label{sec:results}}
In this section, we explore the distribution of sSFH profiles generated by each simulation.  We first examine the regions in UMAP space occupied by each simulation, and then introduce a metric to quantify how many simulations are able to generate a given sSFH.  The combination of these results provides insight into the varying sSFH shapes each simulation is able to produce, introducing a window into the underlying subgrid physics associated with a given sSFH.  

\subsection{Simulation real estate in the UMAP projection space\label{sec:sop}}
\begin{figure*}[hbtp!]
    \centering
    \includegraphics[width=\textwidth]{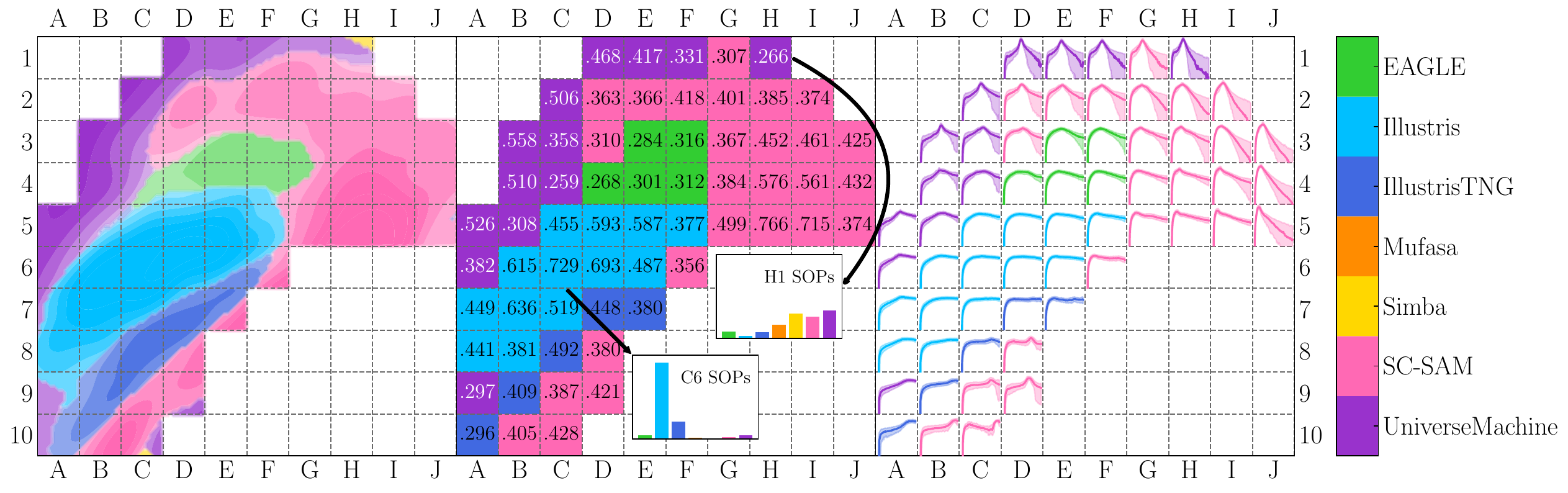}
\caption{Simulation-of-Origin Probabilities (SOPs, $P(\mathrm{sim}|\vec{x})$) -- the conditional probability of any individual simulation producing an sSFH at point $\vec{x}$ in the projection space. Colours follow Figure~\ref{fig:SFMS}, and the layout follows Figure~\ref{fig:kde}. In the \textbf{left} panel, each point is coloured by the simulation with the highest SOP at its location, and filled contours trace increasing SOP values for that simulation, thus providing a map for the regions of the UMAP where each simulation dominates. The \textbf{centre} panel shows the largest cell-averaged-SOP (value printed) and thus provides insight into the distribution amongst simulations in a given cell: high probabilities mean galaxies from one simulation dominate the cell, while smaller probabilities correspond to a more-even spread amongst simulations. Insets show the full SOP distribution across all seven simulations for two example cells: C6 (the densest region of Figure~\ref{fig:kde}) -- strongly dominated by Illustris, and H1 -- an edge cell with no clearly dominant simulation.  The \textbf{right} panel shows the cell-median sSFH coloured by the dominant simulation.}
    \label{fig:sop}
\end{figure*}

\begin{figure*}[hbtp!]
    \centering
    \includegraphics[width=\textwidth]{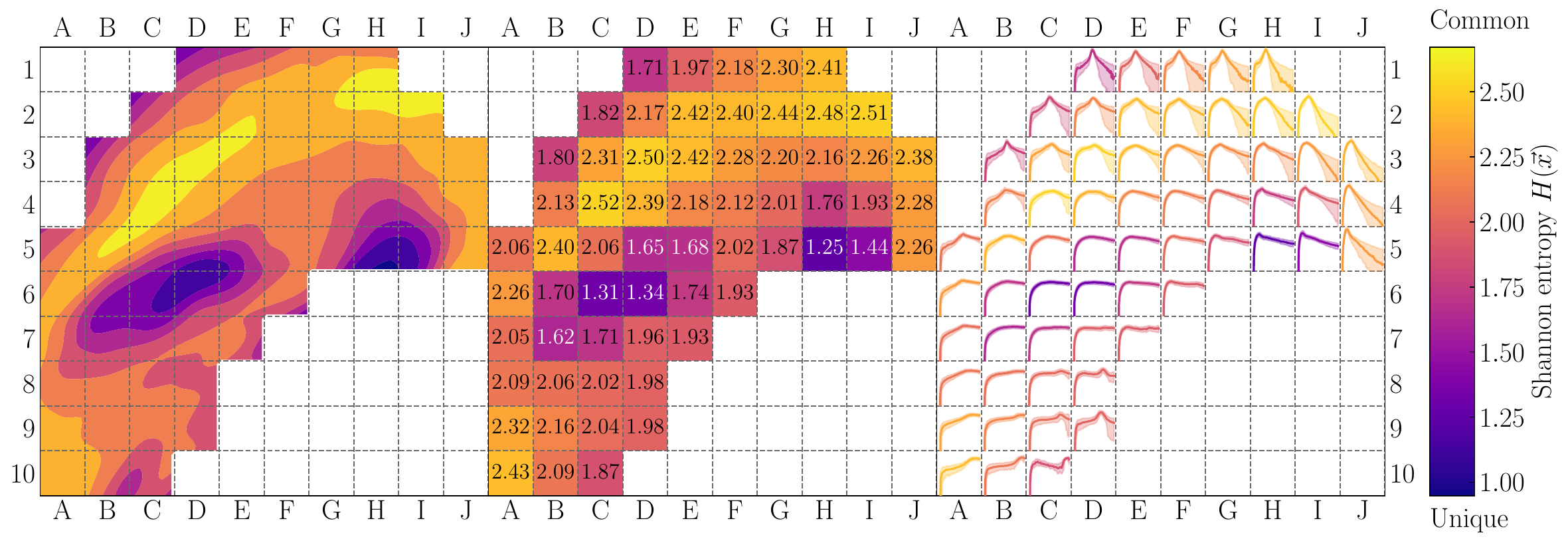}
    \caption{Shannon entropy $H(\vec{x})$ of the SOP distribution across the UMAP: high $H(\vec{x})$ (yellow) highlights regions of \textit{commonality}, corresponding to sSFHs frequently produced by most, if not all, simulations, while low $H(\vec{x})$ (indigo) shows regions of \textit{uniqueness} exclusive to one or two simulations. The layout follows Figure~\ref{fig:kde}, with filled contours in the left panel tracing increasing entropy. A broad ``commonality strip" runs across the top of the taco, covering an arch from cells $\sim$B5 up to G2 and back down to J5; these cells correspond to the lowest maximum SOPs (i.e., fairest spread amongst simulations) in Figure~\ref{fig:sop}.  Likewise, the least common sSFHs amongst the simulations correspond to the largest maximum SOPs (i.e., cells dominated by a single simulation).  The rarest sSFHs live in cells C6 and H5; C6 depicts a steady rise and long plateau, dominated by Illustris, and H5 shows near-immediate assembly followed by slow quenching, primarily produced by SC-SAM.  Other semi-unique patches exist in cells C10 (including the D9 protrusion) -- sSFHs characterized by a double peak, mainly occupied by SC-SAM, and D1 -- steep, central peaks, primarily comprised of UM.  Possible physical prescriptions likely to lead to these trends in commonality and uniqueness are discussed in Section~\ref{sec:discussion}.
    }
    \label{fig:shannon}
\end{figure*}

From Figure~\ref{fig:UMAP}, it is clear that certain simulations or simulation families dominate in some areas of the UMAP space, while other regions are occupied by a mix of most if not all of the simulations. To quantify how likely it is for a given simulation $s$ to occupy a region in UMAP space located at position $\vec{x}$, we introduce the \textit{Simulation-of-Origin Probabilities (SOPs)} $P(s|\vec{x})$. Using Bayes' Theorem,
\begin{align}
    P(s|\vec{x})&=\frac{P(\vec{x}|s)P(s)}{P(\vec{x})}\label{eq:bayes}\\
    &=\frac{P(\vec{x}|s)P(s)}{\sum_i^N P(\vec{x}|s_i)P(s_i)}\label{eq:sop},
\end{align}
where $P(\vec{x})$ has been defined as the normalization by summation across the $N=7$ simulations $s_i$.  

$P(\vec{x}|s)$ in the numerator of Equation~\eqref{eq:bayes} is then given by the KDE over all points in the UMAP space belonging to simulation $s$, calculated as described in Section~\ref{sec:projection} for the total KDE over all simulations.  Since the value of the sum of the total KDE was set to $10^4$, this too must be the value of the total sum over the individual simulation KDEs (i.e., $\sum_i^N P(\vec{x}|s_i)=10^4$).  Thus, each individual KDE is normalized by the proportion of galaxies belonging to the corresponding simulation, such that each KDE is multiplied by a factor of $n_\mathrm{gal}/N_\mathrm{gal}$, where $n_\mathrm{gal}$ is the number of galaxies from simulation $s$ as shown in Table~\ref{tab:sim_specs}, and $N_\mathrm{gal}=63,083$ is the total number of galaxies in our sample.

Finally, we set $P(s)\equiv1/N=1/7$.  Note that each $P(\vec{x}|s)$ has already been implicitly weighted by $n_\mathrm{gal}/N_\mathrm{gal}$ in the KDE normalization described above, so the posterior remains galaxy-count weighted under this equal prior: the two choices -- booking the weighting in the likelihood or in the prior -- yield identical posteriors.  We adopt this convention as the default because it reflects the empirical abundance of each simulation's sample on the UMAP projection, and because, as we show in Appendix~\ref{app:SOP2}, it yields the more accurate per-galaxy classifier.  An alternative construction that strips the galaxy-count weighting -- taking the $P(\vec{x}|s)$ as per-simulation unit-area KDEs while retaining $P(s) = 1/7$ -- is presented as a robustness check there.

Using EAGLE as an illustrative example to plug into Equation~\eqref{eq:bayes}, the probability that a point $\vec{x}$ in UMAP space is occupied by EAGLE is given by 
\begin{align}
    P(\mathrm{EAGLE}|\vec{x})&=\frac{P(\vec{x}|\mathrm{EAGLE})P(\mathrm{EAGLE})}{P(\vec{x})}\\
    \mathrm{SOP_{EAGLE}}&=\frac{\frac{1}{7}\mathrm{KDE_{EAGLE}}}{\frac{1}{7}\sum_i^7 \mathrm{KDE}_{s_i}}.
\end{align}
The SOP is calculated for each of the seven simulations over the same $100\times100$ mesh grid as the evaluated KDE.  Therefore, each point $\vec{x}$ on this $100\times100$ mesh grid is associated with seven SOP values.  The simulation with the highest SOP at point $\vec{x}$ is thus the most likely simulation to occupy that point.  

Figure~\ref{fig:sop} shows the simulation with the highest SOP in the layout of Figure~\ref{fig:kde}.  The left panel shows contours that scale with the SOP value of the most likely simulation over the $100\times100$ mesh grid.  The SOPs are then averaged over each battleship grid cell for all seven simulations, and the simulation with the highest resulting average is shown in the centre and right panels.  We note that Mufasa and Simba have little to no real estate in the projection space: their sSFHs concentrate along the upper crest of the taco and in the D9 protrusion (Figure~\ref{fig:UMAP}), but this territory is shared with UniverseMachine, whose roughly $4\times$ larger galaxy count wins the overlap under the galaxy-count-weighted KDE of Equation~\ref{eq:sop}.  Stripping this weighting returns those cells to Mufasa and Simba (Appendix~\ref{app:SOP2}), confirming that their disappearance from Figure~\ref{fig:sop} is a weighting artefact rather than a shape one.  

Also included in the centre panel are two cutouts showing the SOP distribution across all simulations for a corresponding battleship grid cell.  Although cell H1 is associated with UM, it has one of the lowest SOP values in the entire UMAP space of 0.265, corresponding to a larger spread in likelihood across all simulations.  In cell H1 in particular, there is a nearly equal $\sim25\%$ chance that any given galaxy comes from UM, SC-SAM, or Simba, with the remaining $\sim25\%$ spread amongst the remaining simulations.  On the contrary, cell C6 has one of the highest SOP values of 0.735, meaning that there is nearly a 3/4 chance of any sSFH in that cell coming from Illustris.  This is especially interesting since cell C6 is also the most densely populated cell, as shown in Figure~\ref{fig:kde}.

\subsection{Commonality and uniqueness\label{sec:shannon}}
The SOPs in Figure~\ref{fig:sop} already hint at which sSFH shapes are commonly produced amongst all simulations (cells with a low SOP for the most likely simulation), and which sSFH shapes are rare or even unique to only one or two simulations (cells with a high SOP for the most likely simulation).  To quantify this commonality or uniqueness more precisely, we use the Shannon entropy \citep{Shannon1948}
\begin{equation}
    H(\vec{x})=-\sum_i^N P(s_i|\vec{x})\,\log_2\left(P(s_i|\vec{x})\right).
    \label{eq:shannon}
\end{equation}

Shannon entropy describes the expected level of uncertainty associated with the potential states of a random variable by considering the distribution of probabilities across all potential states.  In this way, it measures the average amount of information in bits needed to describe the state of the variable, such that a higher value for the Shannon entropy -- or more bits required to describe the state -- corresponds to a more uniform spread in probabilities and thus greater uncertainty.

Here, we use Shannon entropy as a metric for commonality versus uniqueness.  Drawing an analogy to thermodynamic entropy, regions in UMAP space with a low SOP for the most likely simulation have a large number of associated microstates; likewise, regions with a high SOP for the most likely simulation have a smaller number of associated microstates.  Thus, Shannon entropy allows us to quantify how common a given sSFH shape is amongst all simulations, with high Shannon entropy corresponding to a more common sSFH that multiple simulations are able to produce, and low Shannon entropy corresponding to a sSFH unique to a small selection of simulations.  

Figure~\ref{fig:shannon} shows the Shannon entropy over UMAP space calculated using the SOPs from Equation~\ref{eq:sop} in the style of Figure~\ref{fig:kde}.  Since the SOPs depend on the individual simulation KDEs, $H(\vec{x})$ is again evaluated over the same $100\times100$ mesh grid, and is plotted in the left panel of Figure~\ref{fig:shannon} as a contour map, with the centre and right panels showing the cell-averaged Shannon entropy.  As expected, the regions with low Shannon entropy align with the regions with a high SOP for the most likely simulation (e.g., cells C6, H5), highlighting sSFHs unique to only one or two simulations.  We also see a ``commonality strip" near the crest of the taco, which again corresponds to cells with low SOPs for the most likely simulation.  We discuss which, if any, subgrid models may be responsible for producing such common or unique sSFH shapes in Section~\ref{sec:discussion}.

\section{Discussion}
\label{sec:discussion}
In this section, we interpret the commonality and uniqueness structure identified in Section~\ref{sec:results} in terms of the physical processes that may be responsible for it. In Section~\ref{sec:explaining-shapes}, we examine the features of the SFH manifold itself, tracing the common shapes to dark matter halo assembly and the unique ones to the subgrid prescriptions of individual simulations. In Section~\ref{sec:dimensionality}, we use the auto-encoder to investigate the intrinsic dimensionality of star formation histories and to explain why it recovers the simulation of origin far more accurately than UMAP. In Section~\ref{sec:MWAs}, we demonstrate an application of our framework by projecting simulated Milky Way analogues into the latent space, and finish in Section~\ref{sec:caveats} with the caveats and limitations that should be considered when interpreting these results.

\subsection{Understanding features in the SFH manifold\label{sec:explaining-shapes}}
As we have seen from Section \ref{sec:results} and previous literature \citep{Iyer2020}, the overall shapes of galaxy SFHs are fairly consistent across different models, rising at early times before peaking and levelling off or falling at late times in a way that can be described with a roughly lognormal \citep{Diemer2017} or double power-law \citep{Iyer2025b} shape on average. However, across different regions of the UMAP space, we find notable deviations from this pattern, along with some models more suited to producing galaxies with certain SFH shapes compared to others. Below, we explore both the commonalities and differences that likely drive these phenomena.

\subsubsection{Convergence from tracing halo assembly\label{sec:conv}}

Across essentially all the models, galaxy SFHs tend to rise and fall, with this behaviour becoming more pronounced at higher stellar masses where quenched fractions are larger. \citet{Iyer2020} found that the in-situ SFHs of central galaxies are coherent with their parent halo accretion histories on long timescales ($\gtrsim 3$ Gyr), suggesting that on these scales the halo growth sets the overall shapes of the SFHs. This makes intuitive sense, since the galaxy's gas supply on long timescales is regulated by its parent halo; it is the basis for ``bathtub''-like models of galaxy formation \citep{Bouche2010, Dave2012, Lilly2013}, in which the star formation rate is set by a quasi-equilibrium between cosmological gas inflow, outflows, and star formation. 

The growth of dark matter halos themselves is a well-understood stochastic process with a characteristic (average) shape set by the hierarchical growth of overdensities in a (mostly) Gaussian random field. Since DM accretion histories have a characteristic rise-and-fall shape\footnote{well described by fitting functions of the form $M_h(z) \sim (1+z)^{af(M_0)}e^{-f(M_0)z}$ \citep{Correa2015}.}, and baryons broadly follow the potential wells of their host halos, galaxy SFHs should inherit this shape on the longest timescales (e.g., \citet{2003MNRAS.341.1253H}). 
The physical machinery modifying the link between star formation rate and dark matter accretion rate is fairly well established; at small halo mass, stellar feedback prevents gas on small scales from turning into stars at the rate gas accretes, but the effect is small for the stellar masses considered here. At large halo mass, a combination of long post-accretion shock gas cooling time \citep{Dekel06, Keres2009, Zinger2018} and AGN feedback prevent accreted gas from turning into stars. The miss-match between stellar and halo mass is largest for these very large halos, loosening the connection between halo assembly and SFR.

To test whether the dominant mode of evolution across the UMAP space is driven by variation in halo mass, we compare the cell-averaged SFHs to two reference curves constructed from the halo-assembly framework (here we drop the ``specific'' \textit{s} as we are comparing only shapes without considering magnitude). The first is the abundance-matched empirical SFH at halo mass $M_h(z=0)$ from \citet{Behroozi_etal_2013}: this is a like-for-like comparison to our cell-averaged SFHs, so fitting each cell's SFH against the Behroozi+13 template library returns an empirical $M_h$ and an associated goodness-of-fit $\chi^2$ per cell.  The resulting empirical $M_h$ is used to generate the ensemble-averaged halo mass accretion history (or dark matter accretion history, DMAH) $\dot{M}_h(t)$, the second curve of comparison, from the \citet{Correa2015} fitting function. The Correa DMAH is \emph{not} an SFH -- it is the growth rate of the dark matter halo itself.  Where the SFH shape tracks the DMAH shape at the same $M_h$, halo accretion alone suffices to set the SFH shape; where it deviates, baryonic physics beyond halo-tracking is shaping the SFH. 

\begin{figure*}
    \centering
    \includegraphics[width=\linewidth]{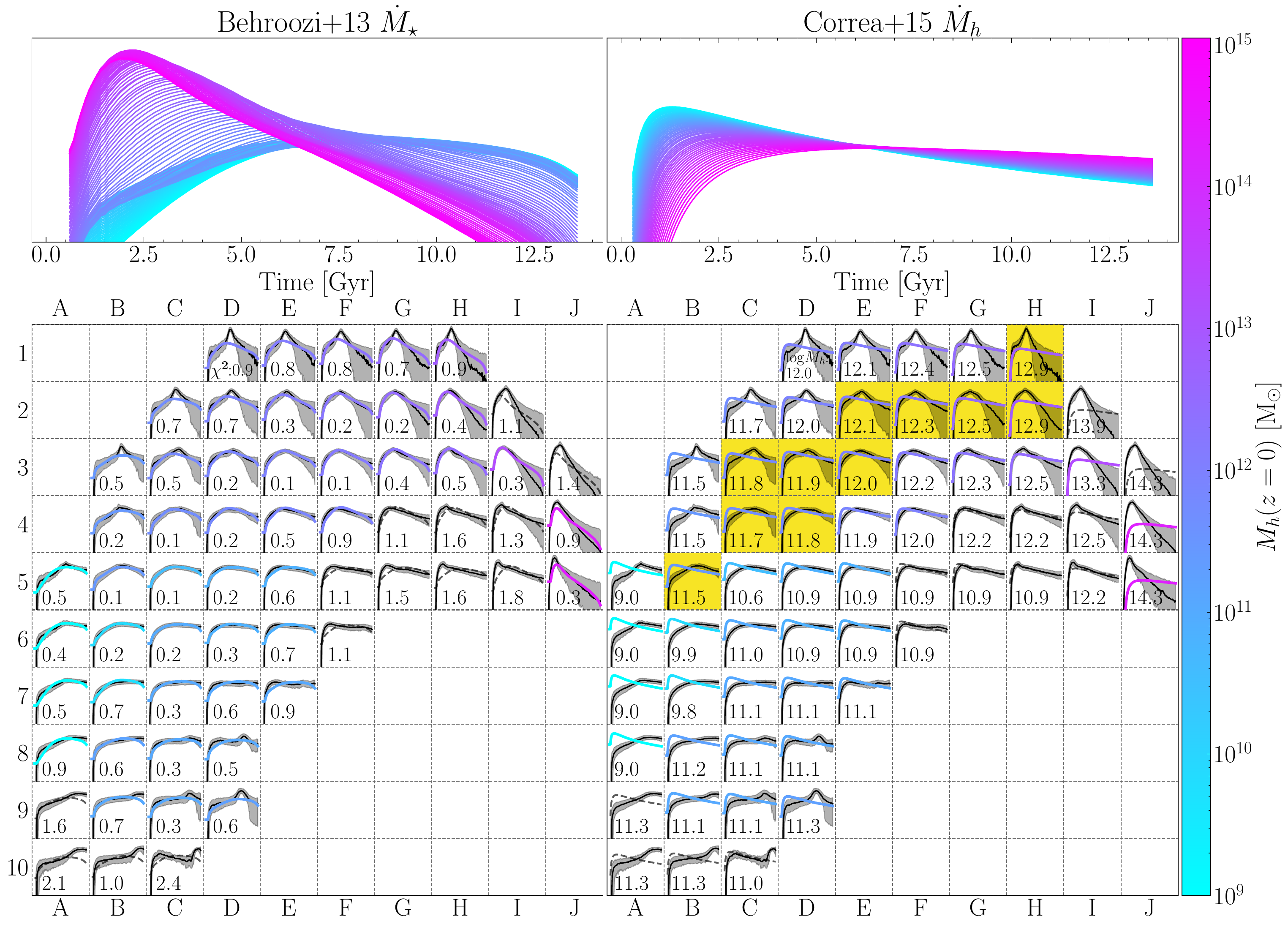}
    \caption{Per-cell comparison of the cell-averaged sSFHs in our $10\times10$ UMAP binning against two assembly references. \textbf{Top:} reference curves for all halo masses, each divided by its median and coloured by $\log M_h$: the \citet{Behroozi_etal_2013} average SFHs (\textbf{left}) and the \citet{Correa2015} dark matter accretion histories $\dot{M}_h(t)$ (\textbf{right}). \textbf{Bottom left:} each cell's median sSFH (solid black, with shaded grey $16$--$84$ percentile band) overplotted with the best-fitting Behroozi SFH at the inferred halo mass $M_h$, rescaled to the cell's amplitude (the fit uses shape only). The Behroozi curve is coloured by $\log M_h$ where $\chi^2 < 1$ for the Behroozi and cell-averaged sSFH fit, and drawn as a dashed grey curve otherwise; each cell is annotated with its $\chi^2$. \textbf{Bottom right:} the same sSFHs overplotted with the Correa DMAH at the same inferred $M_h$, scaled to the sSFH amplitude and drawn in the matching colour where the Behroozi fit is good ($\chi^2 < 1$), as a dashed grey curve elsewhere; each cell is annotated with $\log M_h$. Cells shaded yellow have a high mean Shannon entropy ($\langle H(\vec{x}) \rangle > 2.3$; compare with Figure~\ref{fig:shannon}) and $\chi^2_{\rm SFH} < 1$ to highlight the ``commonality strip". The inferred $M_h$ increases smoothly from the lower left to the upper right of the embedding, from $\log M_h \approx 9$--$11$ to $\approx 13$--$14.3$, consistent with the stellar-mass gradient of Figure~\ref{fig:sm/sfr}; cells at $\log M_h = 9$ sit at the lower edge of the Behroozi grid and should be read as limits. Low-$\chi^2$ cells concentrate in the commonality strip, where the Behroozi SFH at the inferred $M_h$ reproduces the cell's shape well and the Correa DMAH shares its overall rise-and-fall form. Elevated values (up to $\chi^2\approx 2.4$) cluster along the sharply-quenched upper-right edge of the taco and in a few low-mass or double-peaked cells in the lower-left corner, where a smooth reference cannot capture such features. The commonality strip agreement anchors the ``gravitational floor'' argument developed in Section~\ref{sec:discussion}.}
    \label{fig:behroozi_correa_dmtrace}
\end{figure*}

Figure~\ref{fig:behroozi_correa_dmtrace} shows both comparisons cell-by-cell across the UMAP. The inferred $M_h$ in the right panel trends smoothly across the embedding, from $\log M_h \approx 9$--$11$ at the low-mass end to $\log M_h \approx 13$--$14.5$ at the high-mass end, consistent with the stellar-mass gradient visible in Figure~\ref{fig:sm/sfr} and with modest noise in the mid-mass region where the scatter in the stellar mass--halo mass relation is expected to be largest. Across the commonality strip the Behroozi SFHs match the cell-averaged SFHs very well ($\chi^2 \lesssim 0.5$), and the Correa DMAH at the same $M_h$ is also a reasonable shape-level analogue\footnote{We reiterate that here we wish to only compare shapes, and thus have normalized the Correa DMAHs and our galaxy SFHs by their median value for comparison on the same set of axes.} to each cell's SFH (here the commonality strip is highlighted in yellow for cells with an average $H(\vec{x})>2.4$, and neglecting cells with a Behroozi fit $\chi^2<1$). This joint agreement is consistent with the first-order approximation $\mathrm{SFH}(t) \approx f_b\, \epsilon_\star(M_h, z)\, \mathrm{DMAH}(t)$ (e.g., \citet{Mutch2013,RodriguezPuebla_etal_2016,Tacchella2018}, where $\epsilon_\star$ is the instantaneous star-formation efficiency -- the halo accretion history sets the rise-and-fall morphology, and the conversion efficiency is approximately separable from the shape across the mass range probed. 

Deviations from this picture are concentrated in predictable places: the sharply-quenched upper-right regions of the taco show elevated Behroozi $\chi^2$ values and Correa DMAHs that cannot reproduce the late-time truncations, and a sparser set of low-mass and bursty cells -- some of which align with the SC-SAM- and UM-dominated regions analyzed in the next subsection -- show Behroozi failures of a different character, driven by short-lived merger or parametric features that neither a smooth empirical template nor a smooth DMAH curve can encode. The cell-averaged treatment here assumes $\epsilon_\star$ is roughly constant across the relevant mass range; a more refined comparison folding in its mass- and redshift-dependence explicitly, or working at the galaxy level rather than the cell-averaged level, would tighten the quantitative conclusions but is unlikely to change the qualitative picture: the commonality strip is where halo assembly sets the SFH shape, and the deviations are where baryonic physics takes over.

This picture is reinforced by recent results from the CAMELS project \citep{VillaescusaNavarro_2021, VillaescusaNavarro_2023}, where SFHs, feedback strengths, and circumgalactic medium properties can be varied across a broad range of sub-grid parameter values within a single code.  Even when feedback parameters are pushed well outside their fiducial calibrations, the long-timescale shape of the SFH tracks the underlying halo assembly; the scatter shows up primarily at short timescales and in the normalization, not in the overall rise-and-fall morphology \citep{Iyer2025b}. In this sense, the commonality strip seen in the UMAP is not evidence that the simulations agree on galaxy-formation physics -- it reflects a gravitational floor imposed by halo assembly that all simulations have in common. The disagreement then appears in each simulation's models for baryonic feedback, leading to deviation from this DMAH baseline and more unique regions in the UMAP.

\subsubsection{Sharp peaks from mergers and rapid quenching}

There are a few notable areas where the cell-averaged SFHs depart from the Behroozi empirical reference in Figure~\ref{fig:behroozi_correa_dmtrace} (i.e., where the Behroozi SFH is plotted as a grey dashed line rather than in colour), and thus where we would expect baryonic physics to be the primary sculptor of the SFH shape.  These regions are all characterized by a sharp peak in the SFH, and such sharp peaks were established in Figure~\ref{fig:kde} to live on the outskirts of the UMAP.  These ``peaky" regions are separated by the remaining two UMAP ``axes'' -- the timing of the peak and the rate of subsequent quenching -- resulting in four main clusters: a group around H5; a group around C10 (including the D9 protrusion); cells B3, C2, D1, E1, and F1 (which have $\chi^2$ values just below 1); and the top-right edge of the UMAP around cell I2.  These clusters belong to different simulations-of-origin with varying feedback models, and, interestingly, span regions of both commonality and uniqueness as identified in Figure~\ref{fig:shannon}.

The clusters around H5 and C10 both show high uniqueness to SC-SAM.  Here, the sharp peaks likely reflect starbursts triggered when galaxies cross discrete merger or disk-instability thresholds, which drive the SFR up sharply until AGN feedback kicks in and returns the galaxy to a smoother ``disk'' mode \citep{Somerville_etal_2008, Porter_etal_2014, Brennan_etal_2017}. The discrete triggering criterion -- which fires a burst when a merger mass ratio or disk-instability threshold is crossed, rather than letting the local gas physics respond continuously -- is what makes these features so sharp in the SAM relative to the hydrodynamical simulations, whose ISM ``weather'' tends to blur the response out. Depending on when the dominant burst occurs, an SC-SAM galaxy with this behaviour can end up in different cells; it seems the D9 protrusion results from a sharp uptick in star formation just before present day.

Although the group of cells B3, C2, D1, E1, and F1 have fits with $\chi^2\lesssim1$, their values sit on the upper end of this range, and so these galaxies only moderately match the Behroozi SFHs; the sharp peaks are the notable disagreement with the empirical shape.  This region of the UMAP is dominated by UM, and is another region of high $H(\vec{x})$ and thus uniqueness.  UM is an empirical model that parametrizes the SFR as a function of halo mass and redshift with an added stochastic component \citep{Behroozi_2019}, so its sharp peaks are not triggered by discrete physical events as discussed above for the unique SC-SAM regions; they instead reflect features in the $M_\star(M_h, z)$ parametrization and/or the stochastic scatter term. Observationally, mergers can indeed fuel short-timescale starbursts (e.g., \citealp{Asada2024}), so separating SAM-specific burst morphology, UM-specific parametric structure, and genuine merger-driven bursts would require a finer-grained comparison with observationally estimated SFHs of merging systems.

The final sharp-peaked group around cell I2 is characterized by extremely rapid quenching.  Interestingly, this SFH shape is quite common, showing up across all simulations; however, we note that this is also one of the lowest-density regions in the UMAP (c.f. Figure~\ref{fig:kde}) and thus is still quite ``rare" overall.  In other words, many simulations seem to make a few rapid-quenching galaxies.  This abrupt truncation is usually due to strong AGN feedback in most of these models, although individual simulations differ substantially in their feedback models.  
To highlight one example, SIMBA, who lives in this region, uses a two-mode AGN model including a ``jet mode'' active at low Eddington ratios that drives high-velocity bipolar outflows \citep{Dave2019}.  This feedback mode efficiently clears the ISM and produces cliff-like drops in the SFH.  More generally, machine-learning classification applied to EAGLE, TNG, and SIMBA has found that black-hole mass is the dominant predictor of whether a galaxy quenches across all three simulation prescriptions \citep{Piotrowska_2021}.

Direct observational evidence for rapid-quenching modes like these has begun to emerge from JWST. \citet{DEugenio_2024} used NIRSpec IFS of GS-10578 -- a massive post-starburst galaxy at $z \approx 3$ -- to detect fast neutral outflows at $\gtrsim 1000$ km s$^{-1}$ with mass loss rates exceeding the star formation rates, providing direct spectroscopic evidence for ejective AGN feedback in a rapidly quenched system. \citet{Park_2024} identified three distinct quenching trajectories in the Blue Jay survey at $1 < z < 2.5$: a smooth ``old and quiescent'' track, a ``flat + rapid'' decline, and a ``starburst + rapid'' trajectory. The latter two -- both characterized by sharp SFH truncations -- qualitatively match the upper regions of our latent space, suggesting these modes are genuine physical features of the galaxy population rather than artifacts of a specific feedback prescription. The predicted spectroscopic signatures (high $D_n(4000)$ coupled with strong H$\delta_A$) are observationally testable, and their frequency in the observed post-starburst population would constrain the efficiency of ejective AGN feedback directly.

While these rapid-quenching galaxies can provide a valuable test on the strength of feedback needed to match observations, caution must nonetheless be exercised in comparing the two, since observationally reconstructed SED-based methods lose sensitivity for older stellar populations and may smooth out sharp truncations that simulations resolve explicitly.

\subsubsection{A subtle long-timescale signature from Illustris}

There remains one rather distinct region of the UMAP with high uniqueness: the Illustris-dominated cluster centred on C6 (with some TNG contribution) where the cell-averaged SFH has a characteristic shape with a notably narrow spread about the median. A natural first interpretation would be a burstiness signature: Illustris and TNG use different feedback prescriptions from the other hydrodynamical codes, and theoretical work has shown that the PSD of SFR on sub-Gyr timescales is sensitive to feedback coupling and duty cycle \citep{Caplar2019, Iyer2020, Tacchella_2020}. Under that interpretation, Illustris/TNG SFHs could ``flicker'' differently and land the C6 cells in a corner of the map all to themselves through short-timescale texture. Smoothing the SFHs should then blur the signature out.  

Interestingly, rebuilding the UMAP embedding on SFHs that have first been smoothed with a 1 Gyr boxcar produces a near-identical map -- the C6 cluster sits in the same relative location, with the same dominant individual simulations (see Appendix \ref{app:dim_smooth}). However, since UMAP only relies on global, averaged features to classify sSFHs as being similar or dissimilar, we also opt for a more rigorous comparison. 

\begin{figure*}
    \centering
    \includegraphics[width=\linewidth]{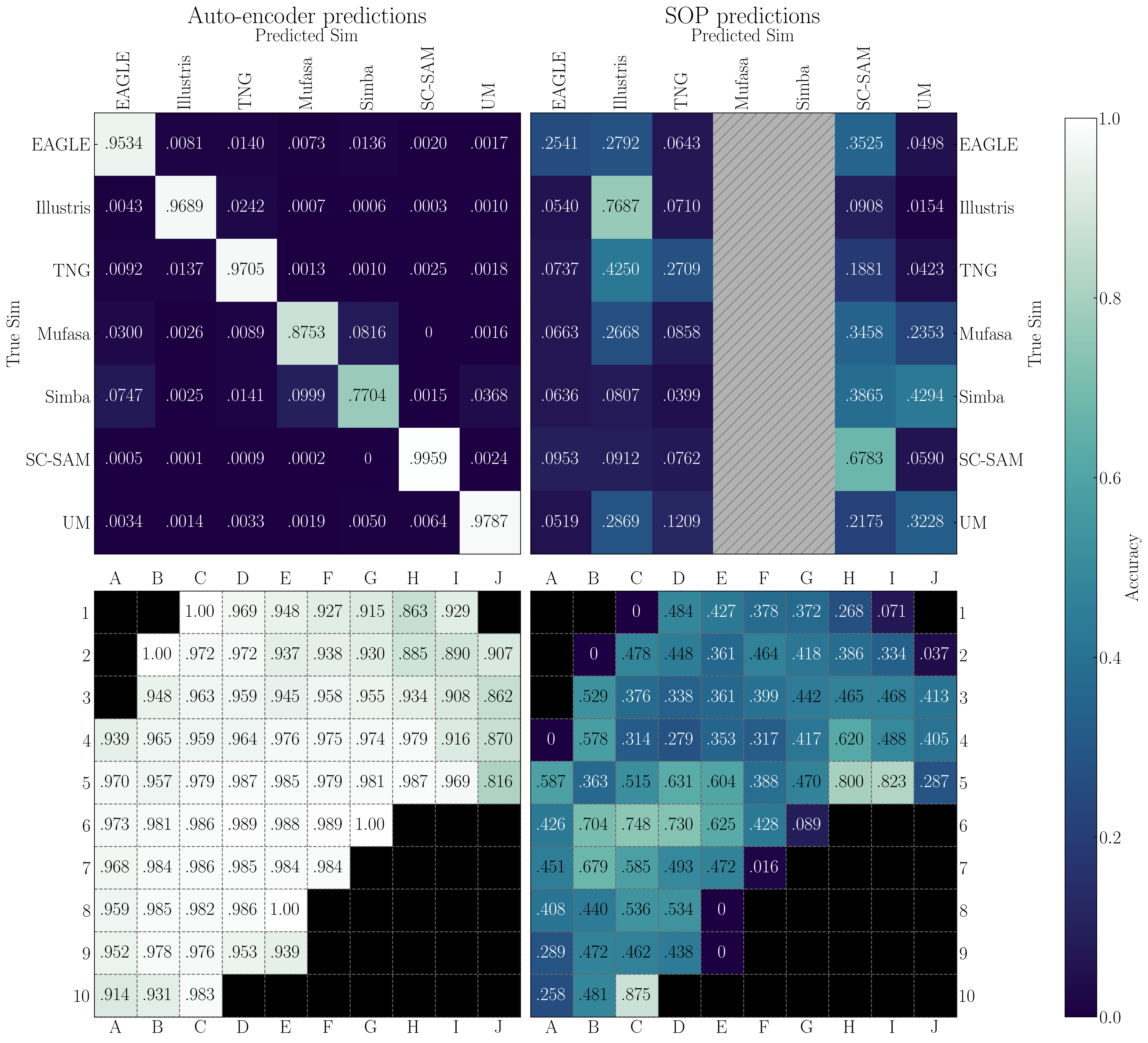}
    \caption{Simulation-of-origin classification accuracy for the multi-objective autoencoder (Appendix~\ref{app:zehao}; \textbf{left column}) and the UMAP-based cell-argmax SOP classifier, which assigns each galaxy to the dominant simulation of its battleship cell (Figure~\ref{fig:sop}; \textbf{right column}). \textbf{Top row:} confusion matrices, with true simulation in rows and predicted simulation in columns (rows normalized). The autoencoder reaches $\sim 95\%$ overall accuracy with a clean diagonal; Mufasa $\leftrightarrow$ Simba is the only substantial off-diagonal confusion. The SOP classifier reaches only $\sim 60$--$70\%$ and cannot predict Mufasa or Simba at all (hashed columns), because neither simulation dominates any battleship cell -- they share their real estate with UniverseMachine, which ``wins'' the overlap in Figure~\ref{fig:sop}. \textbf{Bottom row:} per-cell accuracy on the battleship grid. The autoencoder is uniformly accurate ($\geq 0.82$ in every populated cell), whereas the SOP classifier's per-cell accuracy is structurally bounded by how cleanly a single simulation dominates each cell. The $\sim 25$--$30$ percentage-point gap anchors the argument in Section~\ref{sec:discussion}: simulation-of-origin information in the sSFHs lives in more dimensions than a 2D UMAP can retain.}
    \label{fig:confusion_matrix}
\end{figure*}

As described in Appendix~\ref{app:zehao}, we train a multi-objective auto-encoder (AE) -- a neural network architecture that can perform flexible, non-linear dimensionality reduction -- over the entire suite of simulations with a much larger number of latent dimensions (16 vs UMAP's 2). While we expected the SOP predictions would improve somewhat, we found the AE was able to reach accuracies of as high as $\sim 95\%$ (Figure~\ref{fig:confusion_matrix}), a gap of $\sim 25$--$30$ percentage points over the UMAP embedding (which reaches $\sim 60$--$70\%$). This indicates that the individual sSFHs contain features that can disambiguate between simulations with high confidence, even though in UMAP space the sSFHs look nearly identical. 

To investigate whether this performance was due to short-timescale ($\leq0.1$ Gyr) burstiness, in Appendix~\ref{app:dim_smooth} we also tested whether the accuracy would degrade using the same smoothing strategy applied for our UMAP tests. Again, this only degraded its classification accuracy by $\sim 0.3$--$0.5$ percentage points. This suggests that the sSFH features most useful for distinguishing simulations among our sample are concentrated on $> 1$\,Gyr timescales.

As a result, we suspect Illustris is producing a real, subtle long-timescale shape signature -- a particular balance of early rise, peak time, and late decline -- that the other simulations do not reproduce exactly even though the shapes are extremely similar by-eye. Why Illustris settles into this shape and the other codes do not is not yet clear, but may involve a stable mode of self-regulation in Illustris's feedback model that the other codes' prescriptions cannot sustain, or a level of low-frequency perturbation in the other codes that prevents such a shape from stabilizing.

\subsection{Intrinsic dimensionality of star formation histories\label{sec:dimensionality}}

We now further investigate our interpretation of the gap between UMAP-based and AE-based classification accuracy. As our AE has access to a higher dimensional latent space, we suspect that it should be using them to capture subtle information about the shape of the sSFH in ways that are difficult to capture in just two dimensions (2D). At the same time, we do not expect all dimensions to be equally informative. To determine the intrinsic dimensionality of the 16-dimensional (16D) latent representation our AE uses, we utilize the TwoNN nearest-neighbour estimator \citep{Facco_2017}. As described in Appendix~\ref{app:dim_smooth}, TwoNN estimates imply the raw 136-dimensional sSFH representation has an intrinsic dimensionality of $\sim 24$, the AE's 16D latent space captures $\sim 9$ effective dimensions on our data ($\sim 8$ after smoothing), and the 2D UMAP embedding saturates at its ambient dimensionality of two. 

One interpretation of these findings is that simulations diverge along several independent directions of Gyr-scale SFH morphology, including peak time, rise/fall asymmetry, quench shape, secondary features, and more. As a result, projecting these into two dimensions necessarily collapses most of the separating information. To confirm this, we train multiple independent classifiers (both linear and non-linear) over the 2D UMAP dimensions, the 16D AE latent dimensions, and the 136D raw sSFHs. When the same classifier architecture is trained on each representation, a simple neural network on the 2D UMAP coordinates ultimately caps at $\sim 50\%$ simulation-of-origin accuracy while the same network on the 16D AE latent reaches $\sim 95\%$, nearly identical to our AE's original performance (Appendix~\ref{app:dim_smooth}). This indicates the AE's advantage over UMAP is therefore purely representational capacity for long-timescale morphology, rather than any sensitivity to short-timescale burstiness.

Our results are consistent with independent work on domain generalization across cosmological simulations. \citet{Jo2025} showed that machine-learning models trained on IllustrisTNG fail to generalize to SIMBA and vice versa as the simulation implementations differ systematically along multiple feature directions; our result can be seen as the SFH-shape analogue, in which simulations carry simulation-of-origin information in enough independent directions of SFH morphology that a 2D visualization compresses most of it away. The commonality-vs-uniqueness framing established in the preceding subsections therefore suggests the axis of uniqueness lies on Gyr rather than sub-Gyr timescales. This is observationally encouraging: non-parametric SED-fitting techniques recover Gyr-scale shape features with reasonable fidelity \citep{Pacifici2016, Leja_2019, Carnall_2019, Leja_2019b, Pacifici_2023, Tacchella_2022, Suess_2022}, so the simulation-distinguishing signal should in principle be accessible from observed SFHs -- provided the analysis preserves enough dimensions of variation to resolve it.

We caution, however, that that the exact number of intrinsic dimensions we identify is likely not something fundamental about how galaxy SFHs behave and how galaxies assemble and evolve. As with the subgrid-degeneracy analyses of \citet{VillaescusaNavarro_2021} and \citet{Brown_2024}, this instead likely reflects the number of independent ``subgrid knobs" that differ across the simulations in our suite. These may include, e.g., feedback coupling efficiencies, wind loading factors, AGN mode switches, merger-triggered starburst prescriptions, gas treatment choices, and resolution-dependent stochasticity, which have been compressed by the sample's implicit conditioning on stellar mass and redshift.  The auto-encoder can therefore be better understood to be measuring the dimensionality of the space of implemented models, not of the physical processes those models are intended to represent. The gap between the two is itself one of the quantities that cross-simulation comparisons of this kind appear well-positioned to expose.

\subsection{Does the Milky Way have a unique SFH?\label{sec:MWAs}}
\begin{figure*}[hbtp!]
    \centering
    \includegraphics[width=\textwidth]{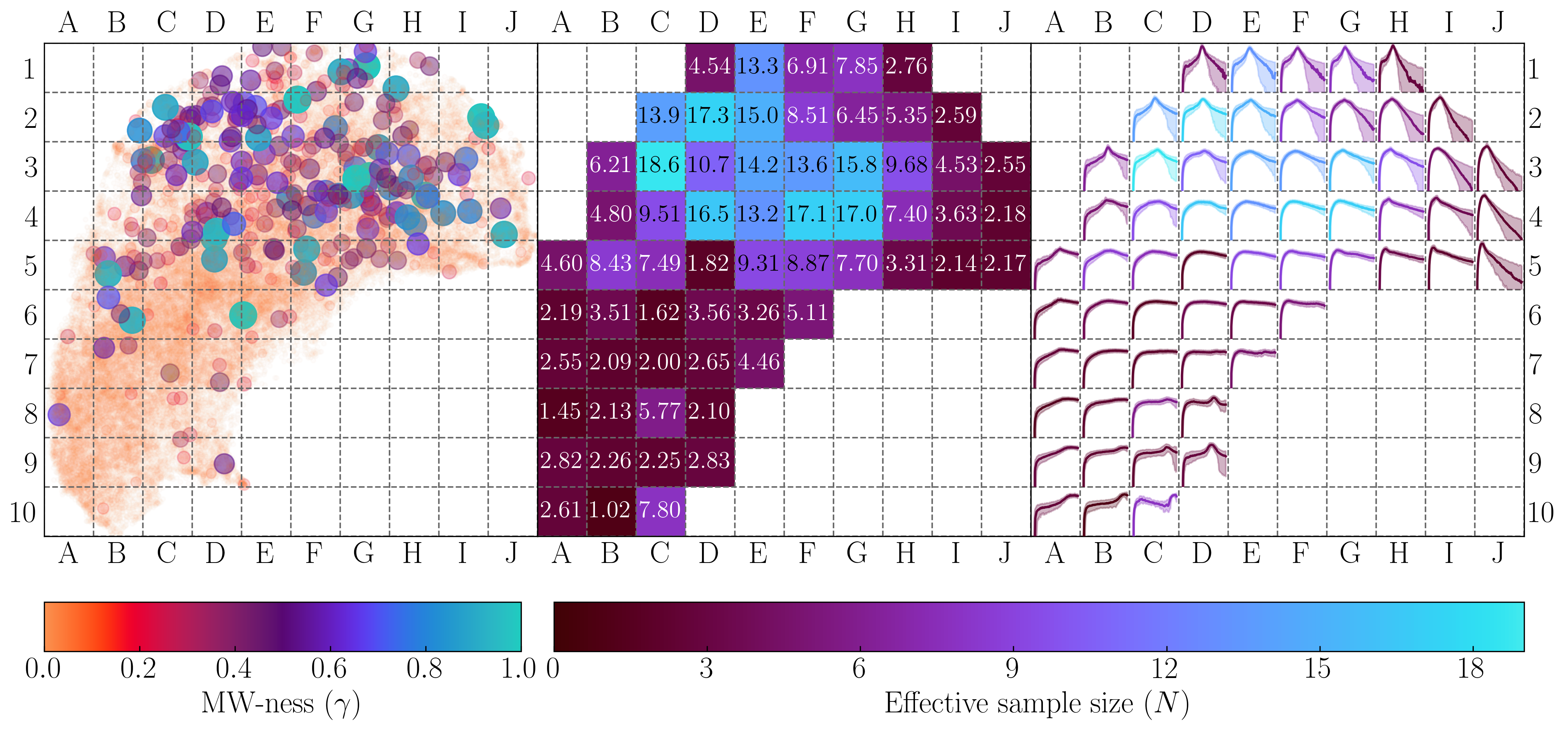}
    \caption{Milky Way analogues (MWAs) projected into the UMAP space, following a variant of the Figure~\ref{fig:kde} three-panel template. MWAs are selected via the $z=0$ $(\SM, \mathrm{SFR})$ criteria of \citet{Savelli2025} (Appendix~\ref{app:MWAs}) and weighted by a continuous ``MW-ness'' score $\gamma \in [0,1]$, with $\gamma=1$ matching the Milky Way's observed values exactly. \textbf{Left:} every galaxy in the UMAP coloured and sized by $\gamma$ (more MW-like galaxies are bluer and have larger points, and galaxies very dissimilar from the MW are faint and orange). \textbf{Centre:} effective sample size $N = \sum_i \gamma_i$ per battleship cell (value printed). \textbf{Right:} cell-median sSFH for each cell, coloured by the cell's $N$. MWAs occupy a broad band of the taco rather than a tight region, avoiding only the extreme low-mass, late-peaking corner (columns A--B, rows 8--10) and the extreme early-quenched corner (columns I--J, rows 1--2). MW-like $z=0$ $(\SM, \mathrm{SFR})$ is therefore reachable via a wide range of SFH morphologies, and the simulation-level spread within this band, rather than the band's existence itself, is what distinguishes simulations at MW mass.}
    \label{fig:MWA}
\end{figure*}

Finally, to demonstrate an application of our findings, we project simulated Milky Way analogues (MWAs) -- selected via the $z=0$ stellar mass and SFR criteria defined in \citet{Savelli2025} (methodological details summarized in our Appendix~\ref{app:MWAs}) -- into our latent space in Figure~\ref{fig:MWA} to determine the cosmological context of Milky Way-like assembly histories.

Direct reconstructions of the Milky Way's SFH from resolved stellar populations now provide a definitive observational benchmark for comparison. \citet{Xiang_2022} used precise ages for $\sim$250{,}000 subgiant stars to reconstruct the MW's formation history, finding that the thick disk formed $\sim$13~Gyr ago -- approximately 2~Gyr before the inner halo assembled via the Gaia-Sausage-Enceladus merger. This reconstruction reveals a relatively smooth, gradually declining SFH for the disk component, with no evidence for the sharp truncation or late-time bursts that characterize specific simulations in our analysis. Subsequent work has refined this picture: \citet{Xiang_2025} identified the ``PanGu'' structure as the oldest surviving disk component at $>$13~Gyr, while \citet{Belokurov_2022} traced distinct chemo-kinematic populations reflecting the transition from early chaotic assembly to ordered disk growth, with \citet{Belokurov_2024} extending this framework to the in-situ-versus-accreted classification of Milky Way globular clusters.

Figure~\ref{fig:MWA} shows that MWAs populate a broad band in of the UMAP rather than a single tight region: while the MWA density is not uniform, it extends across a substantial portion of the manifold and overlaps with the dominant regions of multiple simulations (compare with the SOP map, Figure~\ref{fig:sop}). This is partly by construction: MWAs are selected at fixed $z=0$ $(M_\star, \mathrm{SFR})$, and those quantities correlate with position across the embedding (Figure~\ref{fig:sm/sfr}), so the MWA sample traces a $(M_\star, \mathrm{SFR})$-restricted slice of the SFH manifold. The spread within that slice is what is diagnostic: MW-like stellar mass and sSFR at $z=0$ are reachable via a range of SFH morphologies -- different peak times, different late-time decline profiles, different early-to-late assembly ratios -- and different simulations populate different parts of this reachable set at different densities. 

At the coarsest level, the MWA band sits predominantly within the high-entropy commonality strip identified in Section~\ref{sec:shannon}, so the MW's bulk assembly history -- a gradual build-up followed by a slow decline -- is reproducible across most simulations, and the gravitational-floor interpretation introduced in Section~\ref{sec:conv} applies. The simulation-level spread within the MWA band, however, is the MW-mass cross-section of the broader ``fixed observables, divergent SFH shapes'' phenomenon: \citet{Piotrowska_2021} showed that at fixed black-hole mass the quenching histories still differ systematically between EAGLE, IllustrisTNG, and SIMBA; \citet{Iyer2025b} find the analogous sensitivity in CAMELS, where galaxies with matched present-day properties carry systematically different SFH signatures under different feedback choices; and \citet{Jo2025} showed that machine-learning models trained on TNG fail to generalize to SIMBA even at matched stellar mass. \citet{Savelli2025} themselves performed a targeted TNG-versus-EAGLE comparison of MWA properties in this spirit; extending such pairwise simulation-to-simulation comparisons across our full suite of simulations is a natural next step but lies beyond the scope of this work. 

The MW therefore serves as a useful benchmark: its broad assembly history is ``common'' in the sense of being reproducible across simulations, while its detailed placement within the MWA range -- and the SFH morphology that placement encodes -- is a differentiating constraint on subgrid physics at MW mass.

\subsection{Caveats and limitations\label{sec:caveats}}
Several systematic effects may influence our results and should be considered when interpreting the simulation-specific signatures we identify. We discuss each in turn.

\subsubsection{Resolution effects} The simulations in our sample vary in mass and spatial resolution, with cosmological box simulations (EAGLE, TNG, SIMBA) operating at coarser resolution than zoom simulations (FIRE-2, g14, Marvel/Justice League). We resample all SFHs to a common 0.1~Gyr time resolution and implement stellar-mass cuts ($M_\star > 10^{9}~M_\odot$ for most simulations; $M_\star > 10^{9.5}~M_\odot$ for MUFASA and SIMBA) to mitigate the most direct resolution-driven artifacts. Residual effects may nevertheless persist in the detailed shape of the SFH, since lower-resolution simulations sample star-formation events and short-duty-cycle systems more coarsely than higher-resolution simulations do. Because our main interpretive results rely on Gyr-scale morphology rather than on sub-Gyr texture, these residuals are unlikely to drive the simulation-of-origin signal, but a formal resolution ablation remains a useful future check.

\subsubsection{Sample selection} MUFASA and SIMBA have higher minimum mass thresholds than other simulations due to their resolution limits, which affects the low-mass end of our joint sample. The simulation-level signatures we identify, however, persist when comparisons are restricted to the common mass range ($M_\star > 10^{10}~M_\odot$) where all simulations contribute, so the simulation-of-origin signal is not an artifact of the differing lower mass cuts. The overall shape of the manifold -- particularly in regions dominated by lower-mass galaxies -- does reflect the mass distribution of the joint sample, and comparisons within narrow mass bins (rather than across the full sample) will be necessary to disentangle the mass dependence of simulation-level signatures from the overall simulation-level signatures themselves.

\subsubsection{Simulation box size and environment} The simulation volumes range from $(25 \, h^{-1} \, \mathrm{Mpc})^3$ (Mufasa) to $(100 \, h^{-1} \, \mathrm{Mpc})^3$ (EAGLE, TNG100, Simba), while SC-SAM and UniverseMachine are applied to larger N-body volumes. Smaller boxes undersample rare environments (massive clusters, large voids), which could affect the tails of the SFH distribution. The semi-analytic and empirical models, operating on larger effective volumes, may capture environmental diversity that is truncated in the hydrodynamical runs.

\subsubsection{Coarse time sampling} Our common 0.1~Gyr time grid was set by the coarsest SFH output cadence available across the sample rather than chosen as a tuneable hyperparameter; finer-resolved SFHs do not exist uniformly across all simulations in our sample. The grid is fine enough to resolve the Gyr-scale morphological features that drive our main results (peak time, late-time decline profile, early-to-late assembly ratio, long-timescale undulations), but coarse enough that we make no strong claims about structure at sub-Gyr timescales. The dimensionality analysis in Appendix~\ref{app:dim_smooth} is constructed and interpreted at this same 0.1~Gyr resolution, and the effective dimensionality we report should be read as a statement about long-timescale SFH shape rather than about short-timescale variability. A dedicated analysis using simulation subsets with higher-cadence outputs could extend this to the sub-Gyr regime but lies beyond the scope of the present work.

\subsubsection{Subgrid parameter degeneracies} The CAMELS project has demonstrated that strong degeneracies exist in the mapping from subgrid parameters to observable galaxy properties \citep{VillaescusaNavarro_2021, VillaescusaNavarro_2023}. \citet{Brown_2024} used the ARTEMIS emulator to show that stellar feedback parameters exhibit particularly strong degeneracies -- multiple parameter combinations can produce similar SFH statistics. While our AE successfully distinguishes between the simulations the data originated from, it may not uniquely identify the subgrid parameters responsible for the differences we observe. Future work extending our framework to the full CAMELS parameter space could address this limitation (see also \citealt{Iyer2025b}).

\subsubsection{Physical versus computational dimensionality} The $\sim 8$--$9$ effective dimensions that the AE latent retains on our data (Appendix~\ref{app:dim_smooth}) should not be read literally as the number of independent physical regulators of star formation. A purely physical accounting -- cosmological accretion rate, merger history, gas cooling and feedback coupling, stellar and AGN feedback modes, environmental effects -- is hard to close at that value without double-counting, and a genuinely higher-dimensional physical space would in any case be compressed by our sample's implicit conditioning on mass and redshift. A more natural reading is that the effective dimensionality reflects the number of independent ``subgrid knobs" that differ across the simulations in our suite -- feedback coupling efficiencies, wind loading factors, AGN mode switches, merger-triggered starburst prescriptions, gas treatment choices, and resolution-dependent stochasticity -- rather than the dimensionality of the underlying physics. In this view, the AE is measuring the dimensionality of the space of implemented models rather than of the physical processes those models are intended to represent, and the gap between the two is itself one of the quantities this kind of cross-simulation comparison is positioned to expose.

\subsubsection{Observational validation}
The signatures we identify in Section~\ref{sec:discussion} are, first and foremost, differences between simulations; mapping them onto observed galaxies is a separate and substantially harder problem. Current SED-fitting techniques recover Gyr-scale SFH shape considerably better than sub-Gyr structure \citep{Pacifici2016, Leja_2019}, and the inferred shape depends strongly on the choice of SFH prior \citep{Carnall_2019, Leja_2019b, Pacifici_2023}; \citet{Tacchella_2022} show that even flexible nonparametric parameterizations carry prior-driven biases in the recovered age and decline profile, and \citet{Suess_2022} find that post-starburst ages in particular are underestimated by $\sim 200$~Myr even when the data favour sharp features. The long-timescale morphological features that drive our main results -- peak time, late-time decline profile, early-to-late assembly ratio -- are the features most robustly recoverable from SED fitting, but quantitative comparisons will need to propagate these prior-driven biases into the mapping between simulated and observed SFH distributions rather than comparing the raw distributions directly. Fortunately, JWST has begun to resolve the late-time assembly of individual galaxies directly \citep{Looser2024, Endsley2025, Sun2023}, and targeted high-redshift spectroscopic surveys should increasingly enable distribution-level comparisons against the simulation-level signatures we identify here.

\section{Summary \& Conclusions} \label{sec:conclusions}
Cosmological simulations act as experimental ``universes in a box" to test a variety of physical models in galaxy formation and evolution theory; while the simulations are tuned to reproduce a set of present-day observable scaling relations -- such as the stellar mass--halo mass relation, the galaxy stellar mass function, and the black hole mass--stellar mass relation -- the precise physical models used to evolve the galaxies to these end states vary widely from simulation to simulation.  The combination of individual prescriptions used to model the physics of star formation, stellar feedback, chemical enrichment, black hole seeding and accretion, AGN feedback, and any other physical mechanism taking place below the grid resolution result in set of evolutionary pathways achievable for a given simulation.  In this paper, we compared a sample of $\sim63,000$ SFHs across seven cosmological simulations -- spanning \mbox{(magneto-)}hydrodynamic, semi-analytic, and semi-empirical models -- with the aim of distinguishing which SFH shapes are ubiquitous across several or all simulation families and which are exclusive to only one or two, and subsequently investigating the associated underlying physical models at play.  

Our means of comparison was projecting the cleaned and normalized 136-dimensional sSFHs to a 2D plane via the dimensionality reduction tool UMAP (Figure~\ref{fig:UMAP}).  When splitting the UMAP into 100 battleship-style cells and averaging the sSFHs per cell, we found three implicit axes organizing the galaxies in the UMAP space related to the timing of the peak SFR, the sharpness of the peak, and the rate of quenching (Figure~\ref{fig:kde}, right panel).  While each simulation appears to occupy its own region in the UMAP, there is considerable overlap between these regions.  Thus, we introduced the simulation-of-origin probabilities (SOPs) $P(\mathrm{sim}|\vec{x})$ to determine which simulation was most likely to occupy a given region (Figure~\ref{fig:sop}).  We also use the SOPs to calculate the Shannon entropy at a given point $H(\vec{x})$, with regions of higher Shannon entropy corresponding to sSFH shapes common amongst several or all simulations, and lower Shannon entropy to sSFH shapes unique to only one simulation or simulation family (Figure~\ref{fig:shannon}).  We find one broad ``commonality strip" arching across the top of the UMAP ``taco" surrounded by a number of ``unique islands".  

When fitting the commonality strip SFH shapes against empirical SFHs at a given halo mass $M_h$ from \citet{Behroozi_etal_2013}, we find that the ensemble-averaged DMAH at the same $M_h$ from \citet{Correa2015} fairly describe these SFH shapes to first order (Figure~\ref{fig:behroozi_correa_dmtrace}).  Therefore, the most common SFHs produced amongst our simulation suite trace a ``gravitational floor" driven by cold flows and imposed by hierarchical structure growth, and strong deviations from the DMAHs (i.e., outside the commonality strip) must be caused by individual prescriptions of subgrid baryonic physics which vary between simulations.  There are a few regions in the UMAP where the empirical Behroozi SFH fits are poor ($\chi^2>1$).  One region which cannot be described by the Behroozi SFHs but is still in a common area of the entropy map centres around cell I2, and is characterized by rapid growth followed by steep quenching likely due to strong AGN feedback prescriptions.  

More unique areas of the UMAP entropy map also showcase SFHs which deviate from the empirical Behroozi fits due to very sharp peaks occurring at different stages of evolution, and exist around cells H5 and C10 -- unique to SC-SAM -- and cells B3, C2, D1, E1, and F1 -- unique to UM.  These sharp peaks likely result from starbursts triggered by galaxy mergers in the case of SC-SAM, or the stellar mass parametrization and stochastic scatter term in the case of UM.  One of the most distinct regions of uniqueness lives around cell C6 in an area of the UMAP dominated by Illustris with excellent Behroozi fits ($\chi^2\leq0.3$).  The SFHs in these cells rise and peak early with a slow decline, and exhibit very little scatter about the median.  While we tested short-term burstiness as the cause by smoothing over the SFHs with a 1 Gyr boxcar and discovering no perceptible change in the results and so no significant change, the exact cause of this simple yet unique-to-Illustris shape is not yet understood.  

In the process of analyzing this C6 cluster, we also introduced a multi-objective autoencoder (Appendix~\ref{app:zehao}) which provides a quantitative probe of the same sSFH manifold without restricting the representation to only two dimensions.  While the UMAP embedding achieves $\sim60$--$70\%$ simulation-of-origin classification accuracy, the autoencoder reaches $\sim95\%$, demonstrating that individual sSFHs contain additional information about their simulation of origin that is compressed away in the 2D UMAP representation.  Importantly, smoothing the sSFHs on 1\,Gyr timescales reduces the autoencoder accuracy by only $\sim0.3$--$0.5$ percentage points, indicating that this additional information is primarily encoded in long-timescale SFH morphology rather than short-timescale burstiness.  The autoencoder therefore supports our interpretation of C6 as a subtle long-timescale shape difference, while also demonstrating that simulation-specific differences extend across several independent dimensions of SFH morphology that cannot be visualized simultaneously in two dimensions.  We caution that the resulting $\sim8$--$9$ effective dimensions should not be interpreted as a fundamental dimensionality of galaxy formation, but rather as a measure of the independent directions along which the particular subgrid models represented in our simulation suite differ.

The observational implications of these results are encouraging.  The long-timescale sSFH morphology that distinguishes simulations -- including features such as peak time, decline profile, and early-to-late assembly ratio -- is precisely the regime that current non-parametric SED-fitting methods can recover with reasonable fidelity \citep{Pacifici2016, Leja_2019, Carnall_2019, Pacifici_2023, Tacchella_2022, Suess_2022}.  Thus, if the differences between simulations arise primarily through long-timescale features in galaxy SFHs, as our results suggest, they should in principle be accessible from observations rather than requiring diagnostics of sub-Gyr variability.  Quantitatively connecting these reconstructed SFHs to the simulation distributions, while accounting for the prior-driven biases inherent to SED fitting, is an important direction for future work.

As a brief demonstration of this application, we also projected simulated Milky Way analogues (MWAs) onto the simulation-derived UMAP.  MWAs occupy a broad strip of the manifold, predominantly within the commonality strip, rather than a single characteristic SFH region.  This indicates that the broad, gradually declining assembly history of the Milky Way is reproducible across simulations, while the substantial variation in SFH morphology among MWAs provides a potential additional constraint on the physical prescriptions governing galaxy evolution at Milky Way mass.

More broadly, our results demonstrate that galaxies with similar present-day properties can retain substantial differences in their evolutionary histories as a consequence of the physical prescriptions used to model their unresolved baryonic physics.  The commonality strip suggests that a substantial fraction of SFH diversity is set by the gravitational assembly of structure, while the unique regions reveal departures introduced by the different subgrid models implemented in each simulation.  These differences persist even among galaxies selected to have comparable present-day properties, highlighting that agreement with $z=0$ scaling relations does not imply agreement in the underlying evolutionary pathways.  Extending this analysis to larger and more diverse simulation suites will test the robustness of the commonality--uniqueness structure and determine how the dimensionality of simulation-specific SFH variation changes as additional physical prescriptions are introduced.  At the observational end, distribution-level comparisons with reconstructed galaxy SFHs, while accounting for the uncertainties and prior-driven biases of SED fitting, offer a promising route to testing whether these simulation-specific differences are present in the real galaxy population and potentially breaking degeneracies between models of galaxy formation.

\begin{acknowledgments}
The authors would like to thank Rachel Somerville for insightful conversations and Bart Ripperda for helpful advice. JSS would like to thank Rebecca Bleich for her support.

AMS and NM are supported in part by the Natural Sciences \& Engineering Research Council of Canada (NSERC) [funding reference numbers 568580, RGPIN-2023-04844, and RGPIN-2023-04901], and the Canadian Space Agency (23JWGO2A01).  AMS acknowledges the support of NSERC [funding reference numbers 585825-2023, 589240-2024].  Cette recherche a \'et\'e financ\'ee par le Conseil de recherches en sciences naturelles et en g\'enie du Canada (CRSNG), [num\'eros de r\'ef\'erence 568580, RGPIN-2023-04901, 585825-2023, 589240-2024]. AMS was also supported by the Walter C. Sumner Memorial Fellowship and the Mary and Ron Martin Graduate Fellowship in Astrophysics, with additional support by a grant from the Simons Foundation (MP-SCMPS-00001470). JSS and MAK were also supported by NSERC Discovery Grant RGPIN-2023-04849 and a University of Toronto Connaught New Researcher Award.  Support for KI was provided by NASA through the NASA Hubble Fellowship grant HST-HF2-51508 awarded by the Space Telescope Science Institute, which is operated by the Association of Universities for Research in Astronomy, Inc., for NASA, under contract NAS5-26555.  EA was supported by the Dunlap Institute for Astronomy \& Astrophysics, which is funded through an endowment established by the David Dunlap family and the University of Toronto. ZP and CK acknowledge the Research Opportunities Program at the University of Toronto.  

AI was lightly used while preparing the manuscript. Claude was used to set up and run some of the autoencoder experiments and help summarize their results, and Claude along with Gemini for editing portions of the manuscript text.
\end{acknowledgments}


\software{numpy~\citep{numpy}, matplotlib~\citep{matplotlib}, cmasher~\citep{cmr},
            scipy~\citep{scipy}, UMAP~\citep{McInnes2020umap}, scikit-learn~\citep{sklearn}}

\appendix

\section{Zoom Simulations} \label{app:zoom}
\begin{figure*}[hbtp!]
    \centering
    \includegraphics[width=0.49\textwidth]{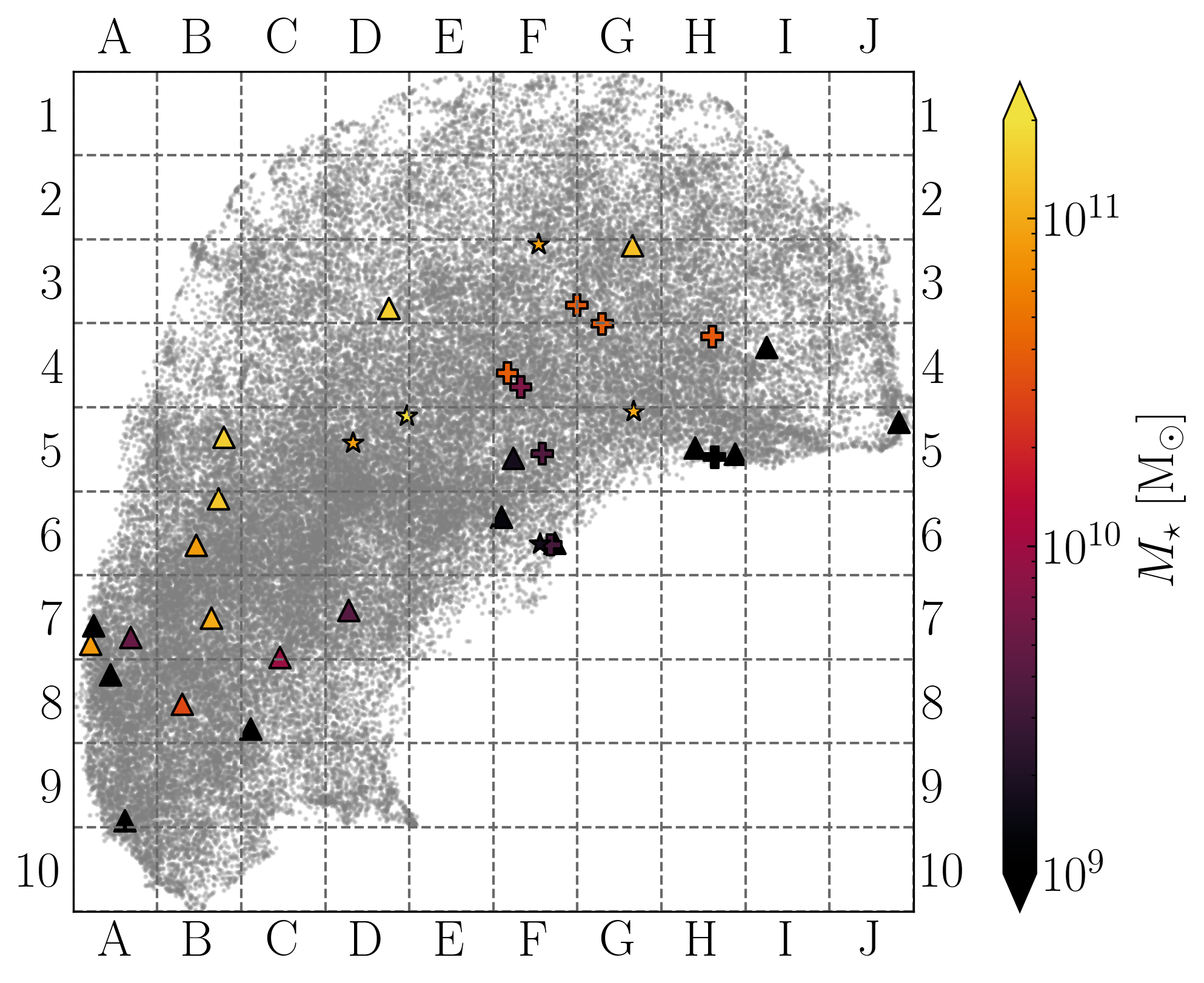}
    \includegraphics[width=0.49\textwidth]{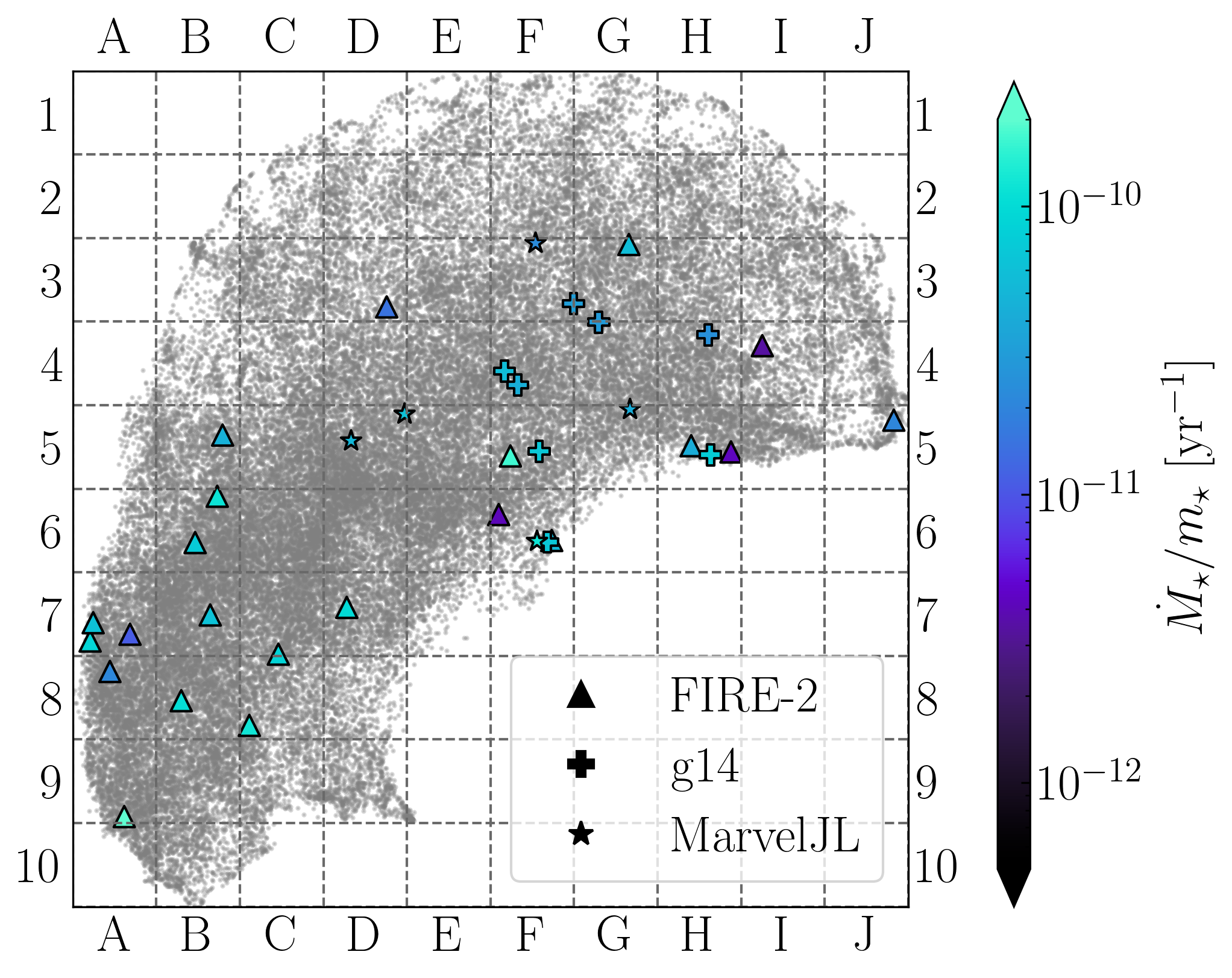}
    \caption{Zoom SIM SFMS overtop greyscale SFMS from Figure \ref{fig:SFMS}.}
    \label{fig:zoom}
    \end{figure*}
Included in our sample of available data from ~\citet{Iyer2020} are a handful of galaxies from zoom simulations FIRE-2 \citep{Hopkins_2018}, g14 \citep{Governato_etal_2012, Munshi_etal_2013, Brooks_Zolotov_2014, Christensen_etal_2016, Brooks_Christensen_2016, Brooks_etal_2017, Christensen_etal_2018}, and Marvel/Justice League \citep[MarvelJL;][]{Bellovary_2018}.  Zoom simulations are a valuable asset to studies of galaxy evolution as they are able to simulate an individual galaxy or halo at much higher resolution with more physically motivated prescriptions.  We neglected to use the zoom simulations in our main analysis as the sample size is too small to perform a meaningful population analysis.  Instead, we have ``painted" the zoom simulations onto our existing UMAP projection using the \texttt{reducer.transform} feature.  

Figure~\ref{fig:zoom} is a repeat of Figure~\ref{fig:sm/sfr} with the addition of the zoom simulations against the grey background of our main galaxy sample.  The zoom simulations do not seem to follow the trends in stellar mass and sSFR observed in the main galaxy sample, which is not unexpected since the sample size is so small and these sSFHs were not included in the UMAP training.  

What is more interesting is the spread of points from each simulation.  FIRE-2 contributes 22 galaxies with the widest spread across UMAP space clustered into three main groups: 1) the left side of the taco in columns A, B and C, corresponding to sSFHs displaying slow assembly that has not yet turned over to quenching; 2) cells F5 and F6, a region with rapid assembly, an early peak, and slow quenching; and 3) cells H5 and I4, with very early and sharply peaked sSFHs.  The 8 galaxies from g14 on the other hand is mainly contained in the perimeter around cells F4 and G4, showing rapidly assembling and moderate quenching.  MarvelJL, with only 5 galaxies, is relatively spread out with no clear clustering.   

\section{Alternative simulation of origin probabilities\label{app:SOP2}}

The simulation-of-origin probabilities defined in Section~\ref{sec:sop} carry an implicit galaxy-count weighting.  Because the per-simulation KDEs in Equation~\ref{eq:bayes} are normalized so that their sum integrates to $10^4$ on the $100\times100$ mesh, each is effectively rescaled by its share of the total sample, $n_\mathrm{gal}/N_\mathrm{gal}$. The map in Figure~\ref{fig:sop} reflects both the shape of each simulation's sSFH distribution and its sample size.  Mufasa and Simba, with $\lesssim 2{,}000$ galaxies each against UM's $7{,}361$, therefore lose every cell where their distribution overlaps with UM purely due to the latter's increased sample size, regardless of how distinctive the underlying shapes might be.

\begin{figure*}
    \centering
    \includegraphics[width=\linewidth]{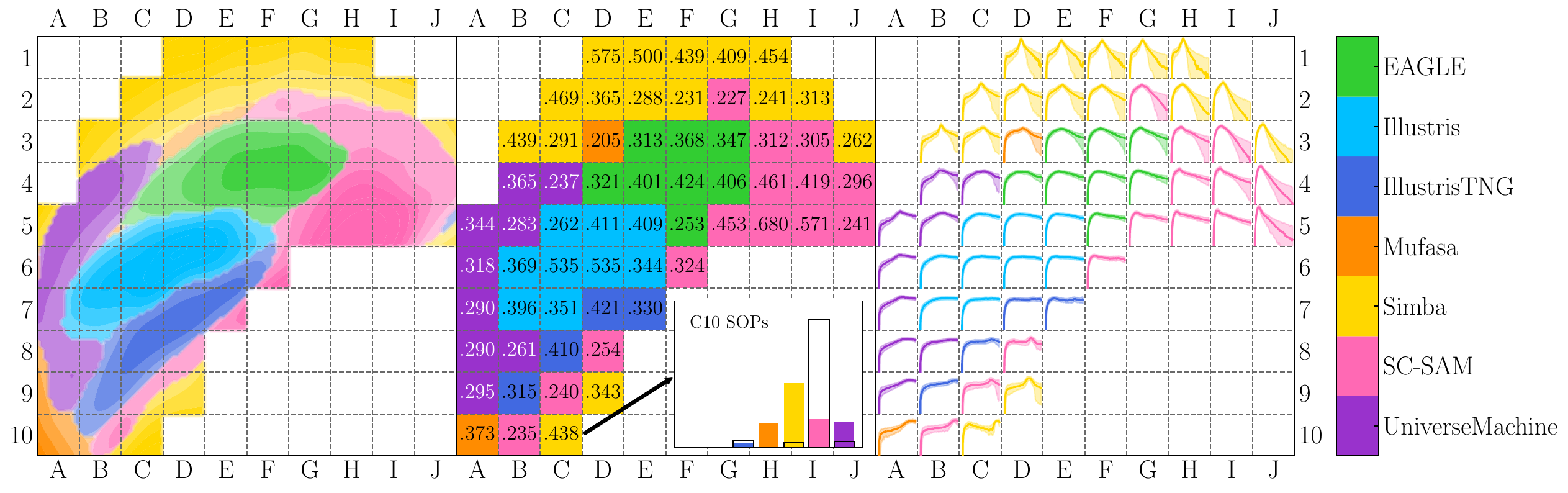}
    \caption{Simulation-of-origin probabilities under the unweighted (equal-weighting) variant described in Appendix~\ref{app:SOP2} -- the analogue of Figure~\ref{fig:sop} with the galaxy-count weighting removed from $P(\vec{x}|s)$. The layout follows Figure~\ref{fig:sop}: \textbf{left}, each mesh point coloured by the dominant simulation with filled contours tracing the dominant SOP value; \textbf{centre}, cell-averaged dominant SOP on the battleship grid (value printed); \textbf{right}, cell-median sSFH coloured by the dominant simulation. Mufasa and Simba now claim cells along the upper crest of the taco and through the D9 protrusion -- real estate they shared with UniverseMachine in Figure~\ref{fig:UMAP} but ceded to it under the weighted prior of Figure~\ref{fig:sop}. Illustris/TNG still rule the C6 cluster, SC-SAM still dominates the upper-right quenched corner, and EAGLE's foothold in the central taco is unchanged; the overall topology of the map is therefore largely preserved, but the small-sample simulations are no longer structurally zeroed out.}
    \label{fig:sop2}
\end{figure*}

We therefore run a complementary analysis where we recompute the SOPs with the galaxy-count weighting removed. The individual $P(\vec{x}|s)$ are taken as the raw (per-simulation unit-area) KDEs, and we assigned an equal prior $P(s) = 1/7$ to each simulation while keeping the remaining analysis assumptions fixed. Figure~\ref{fig:sop2} shows the dominant per-simulation coverage in the (unmasked portion of the) $100\times100$ meshgrid SOP map shifts as expected: the small-sample simulations gain ground (Mufasa $0\% \to 4.24\%$, Simba $0.35\% \to 24.42\%$, EAGLE $6.73\%\to13.84\%$) while the large-sample codes lose it (Illustris $22.48\% \to 13.42\%$, SC-SAM $40.40\% \to 22.23\%$, UM $19.73\% \to 11.52\%$). As IllustrisTNG sits near the crossover, it change only modestly ($10.31\%\to10.34\%$).  In the battleship view, Mufasa and Simba now ``claim'' cells along the upper crest of the taco and through the D9 protrusion -- precisely the regions where they coexisted with UM in Figure~\ref{fig:UMAP} but were outvoted under the weighted prior.  The dominant-simulation picture elsewhere is largely preserved: Illustris/TNG still rule the C6 cluster, SC-SAM still dominates the upper-right quenched corner, and EAGLE's foothold in the central taco is unchanged.

\begin{figure*}
    \centering
    \includegraphics[width=\linewidth]{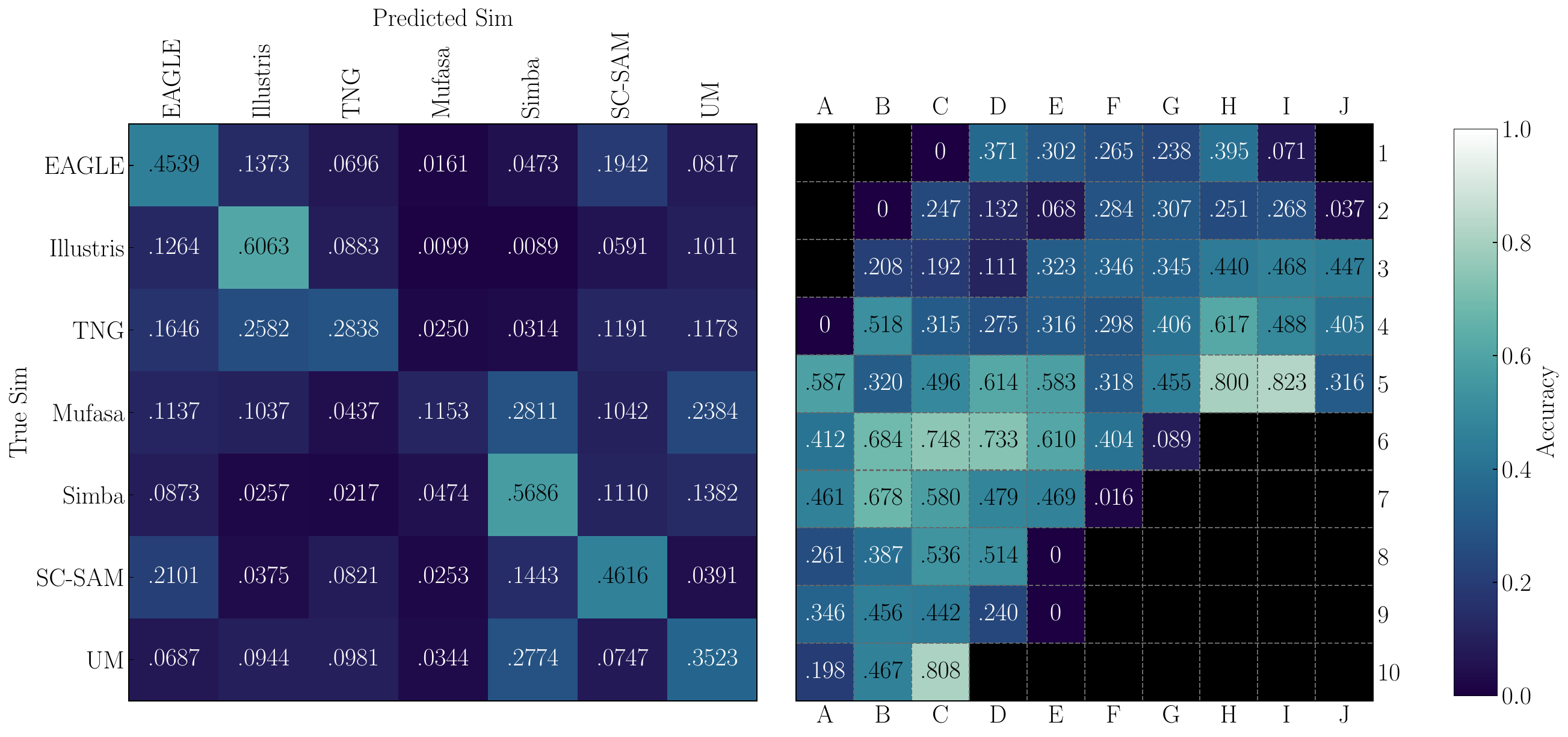}
    \caption{Classification accuracy under the unweighted SOP variant of Figure~\ref{fig:sop2} -- the analogue of Figure~\ref{fig:confusion_matrix} for the equal-weighting alternative discussed in Appendix~\ref{app:SOP2}. The layout follows Figure~\ref{fig:confusion_matrix}: \textbf{left}, confusion matrix for the UMAP-based cell-argmax alternative-SOP classifier (right), with true simulation in rows and predicted simulation in columns (rows normalised); \textbf{right}, per-cell accuracy with the alternative-SOPs on the battleship grid. The Mufasa and Simba columns of the SOP matrix are no longer structurally empty -- both simulations now claim cells and so register non-zero recall -- but off-diagonal mass accumulates in those columns at the expense of the UniverseMachine diagonal. Per-galaxy classification accuracy falls from $42\%$ (weighted prior, Figure~\ref{fig:confusion_matrix}) to $37\%$ (unweighted prior, this figure); the gain in minority-class coverage is outweighed by the re-labelling of UniverseMachine galaxies in the cells Mufasa and Simba newly dominate.}
    \label{fig:accuracy2}
\end{figure*}

The redistribution does not, however, translate into a more accurate classifier.  Under the per-galaxy $\arg\max_s P(s|\vec{x})$ assignment used for Figure~\ref{fig:confusion_matrix}, classification accuracy drops from $42\%$ (weighted prior) to $37\%$ (unweighted prior).  The UM galaxies that populate the cells Mufasa or Simba now claim flip from correct to incorrect predictions, and the gain in Mufasa/Simba recall is too small to compensate.  The confusion matrix in Figure~\ref{fig:accuracy2} shows this directly: the Mufasa and Simba columns are no longer structurally empty, but off-diagonal mass accumulates in them at the expense of the UM diagonal.

As a result, we therefore opt to keep unweighted variant as a robustness check rather than a replacement for the analysis in the main text.  It does, however, confirm that Mufasa's and Simba's sSFH shapes are distinctive enough to surface once the galaxy-count weighting is stripped out, offering an independent view of the simulation-level signatures mapped in Figure~\ref{fig:UMAP}.


\section{Multi-objective auto-encoder\label{app:zehao}}

\subsection{Motivation}

The UMAP embedding in Figure~\ref{fig:UMAP} is a two-dimensional projection, and the cell-level analyses in Sections~\ref{sec:projection}--\ref{sec:shannon} inherit its dimensionality.  As a complementary probe of how much simulation-specific information the sSFHs carry, we train a convolutional auto-encoder with a shared low-dimensional bottleneck and three task-specific decoder heads: (i) sSFH reconstruction, (ii) regression onto the present-day stellar mass and star formation rate, and (iii) classification of the simulation of origin.  The multi-task formulation acts as an inductive bias \citep{Caruna1997}: the bottleneck must encode enough information to identify the simulation and reproduce integrated $z=0$ properties, not just to compress the sSFH.  In combination with UMAP, the auto-encoder provides a quantitative check on the visual impression built up in Section~\ref{sec:results} related to how diagnostic sSFH shape actually is of the underlying subgrid physics.

\subsection{Data preprocessing\label{app:zehao:data}}

The input sample is the same $63{,}083$ galaxies used in the UMAP analysis, drawn from the seven cosmological simulations in the \citet{Iyer2020} compilation and subject to the mass cuts described in Section~\ref{sec:data}. Each sSFH is interpolated onto the same $136$-bin grid spanning $0.1$--$13.6~\mathrm{Gyr}$ in steps of $\Delta t = 0.1~\mathrm{Gyr}$ used throughout the main text.  To isolate sSFH \emph{shape} from total stellar mass, we again area-normalise each history,
\begin{equation}
\widetilde{\mathrm{sSFH}}(t) = 500\,\frac{\mathrm{sSFH}(t)}{\int \mathrm{sSFH}(t')\,\mathrm{d}t'},
\label{eq:ae_norm}
\end{equation}
where the integral is evaluated via the trapezoidal rule at the native grid spacing.  The target area of $500$ is arbitrary -- it simply sets the numerical range in which the network operates such that the values are of order unity -- and is held fixed across the sample.  Two auxiliary targets are extracted directly from the simulation catalogues: $\log_{10}(\SM/\MSun)$ and $\mathrm{arcsinh}(\mathrm{SFR})$.  The $\mathrm{arcsinh}$ transform handles the SFR dynamic range and the small fraction of non-positive SFR values without requiring an offset or floor.

\subsection{Architecture\label{app:zehao:arch}}

The encoder is a stack of four $1$D convolutional blocks (each a Conv1d layer followed by ReLU and stride-$2$ max-pooling), followed by three fully connected layers that collapse the flattened feature map to a $16$-dimensional latent vector.  All convolutions use $16$ filters; successive kernel sizes ($4, 4, 5, 5$) and the four rounds of pooling are chosen so the final feature tensor has shape $(16,6)=96$, matching the flatten dimension.  The latent layer carries no activation, allowing each coordinate to span the real line.

Three decoder pathways act in parallel on the latent vector.  The \emph{reconstruction} pathway mirrors the encoder with transposed convolutions back to the original $136$-bin grid.  The \emph{classification} pathway is a three-layer fully connected network mapping the latent to $7$ raw logits (one per simulation), with softmax applied inside the cross-entropy loss rather than as an explicit output non-linearity.  The \emph{regression} pathway is a four-layer fully connected network with GELU activations \citep{Hendrycks2016} mapping the latent to the two auxiliary targets.  The full layer configuration is given in Table~\ref{tab:ae_arch}.

\begin{table*}[t]
\centering
\caption{Layer-by-layer configuration of the multi-objective auto-encoder.  All Conv1d and ConvTranspose1d layers use $16$ filters; max-pooling is applied with stride $2$ after each encoder convolution.  The three decoder pathways share the encoder and latent bottleneck.}
\label{tab:ae_arch}
\begin{tabular}{llll}
\hline
Stage & Layer(s) & Configuration & Output shape \\
\hline
Encoder input & --- & sSFH on $136$-bin grid & $(1, 136)$ \\
 & Conv1d$_1$ + ReLU + MaxPool & kernel $4$, padding $1$ & $(16, 67)$ \\
 & Conv1d$_2$ + ReLU + MaxPool & kernel $4$, padding $1$ & $(16, 33)$ \\
 & Conv1d$_3$ + ReLU + MaxPool & kernel $5$, padding $1$ & $(16, 15)$ \\
 & Conv1d$_4$ + ReLU + MaxPool & kernel $5$, padding $1$ & $(16, 6)$ \\
 & Flatten + FC$_{96\to 64\to 32\to 16}$ & ReLU on hidden layers & $(16)$ \\
\hline
Reconstruction head & FC$_{16\to 32\to 64\to 96}$ + reshape & ReLU on hidden layers & $(16, 6)$ \\
 & ConvTranspose1d$_1$ + ReLU & kernel $5$, stride $2$ & $(16, 15)$ \\
 & ConvTranspose1d$_2$ + ReLU & kernel $4$, stride $2$ & $(16, 32)$ \\
 & ConvTranspose1d$_3$ + ReLU & kernel $5$, stride $2$ & $(16, 67)$ \\
 & ConvTranspose1d$_4$ & kernel $4$, stride $2$ & $(1, 136)$ \\
\hline
Classification head & FC$_{16\to 100\to 100\to 7}$ & ReLU on hidden layers & $(7)$ \\
\hline
Regression head & FC$_{16\to 200\to 200\to 200\to 2}$ & GELU on hidden layers & $(2)$ \\
\hline
\end{tabular}
\end{table*}

\subsection{Loss function\label{app:zehao:loss}}

The network is trained with a weighted sum of the three task-specific terms,
\begin{equation}
\mathcal{L}_{\rm total} = w_{\rm sfh}\,\mathcal{L}_{\rm sfh} + w_{\rm props}\,\mathcal{L}_{\rm props} + w_{\rm sim}\,\mathcal{L}_{\rm sim},
\label{eq:ae_loss}
\end{equation}
with $\mathcal{L}_{\rm sfh}$ and $\mathcal{L}_{\rm props}$ the mean-squared-error losses on the sSFH reconstruction and the two auxiliary regression targets, and $\mathcal{L}_{\rm sim}$ the cross-entropy between the $7$-way classification logits and the one-hot simulation label.  We adopt task weights $w_{\rm sfh} = 0.01$, $w_{\rm props} = 10$, and $w_{\rm sim} = 1$.  The small reconstruction weight compensates for the large numerical range of the area-normalised sSFHs (Eq.~\ref{eq:ae_norm}, target area $500$), which would otherwise dominate the gradient; the large regression weight emphasises recovery of the integrated $z=0$ quantities, which both anchors the latent space to physically meaningful coordinates and serves as a built-in sanity check against the simulation catalogues.

The training sample is substantially imbalanced, by nearly an order of magnitude between the smallest (Mufasa) and largest (Illustris) classes.  To ensure the minority codes are not overshadowed, we apply inverse-frequency class weights
\begin{equation}
w_{\rm class}(k) = \frac{N_{\rm total}}{N_k}
\label{eq:ae_classweights}
\end{equation}
to the cross-entropy term (mirroring Appendix \ref{app:SOP2}).  The weighted variant is the one reported throughout this appendix, and is also the configuration underlying all main-text confusion-matrix panels (Figure~\ref{fig:confusion_matrix}); an unweighted variant trained in parallel as a sensitivity check yields qualitatively similar classification performance.

\subsection{Training procedure\label{app:zehao:train}}

Training uses $5$-fold cross-validation: the galaxy sample is partitioned into five disjoint folds, and for each fold a fresh model is trained on $80\%$ of the data and evaluated on the held-out $20\%$.  All predictions reported in the paper are therefore \emph{out-of-sample}, with every galaxy scored by a model that never saw it during training.  Within each fold the network is optimised with Adam \citep{Kingma2015} under a one-cycle cosine-annealing learning-rate schedule \citep{Smith2018}, peaking at $3\times 10^{-3}$ after a $5\%$ warmup and decaying to three orders of magnitude below the peak.  Gradient clipping to an $\ell_2$ norm of unity stabilises the warmup phase.  Each fold is trained for $300$ epochs with a batch size of $512$ (the unweighted sensitivity run uses $1024$, but we found performance insensitive to these choises of batch sizes).  Table~\ref{tab:ae_train} summarises the hyperparameters.

\begin{table}[t]
\centering
\caption{Training hyperparameters for the multi-objective auto-encoder.  The weighted (main) and unweighted (sensitivity) runs share everything except the batch size and the inverse-frequency class weights applied to $\mathcal{L}_{\rm sim}$.}
\label{tab:ae_train}
\begin{tabular}{ll}
\hline
Parameter & Value \\
\hline
Optimiser & Adam \\
LR schedule & OneCycleLR (cosine) \\
Maximum LR & $3\times 10^{-3}$ \\
Warmup fraction & $5\%$ of steps \\
Final LR & $\sim\! 3\times 10^{-7}$ \\
Gradient clipping ($\ell_2$) & $1.0$ \\
Epochs per fold & $300$ \\
Batch size (weighted/unweighted) & $512\,/\,1024$ \\
Cross-validation folds & $5$ \\
\hline
\end{tabular}
\end{table}

\subsection{Performance\label{app:zehao:perf}}

Aggregated across the five folds, the weighted variant reaches an out-of-sample classification accuracy of $95.5\%$. For the auxiliary regression targets, out-of-sample root-mean-squared errors are $0.25~\mathrm{dex}$ on $\log_{10}(\SM/\MSun)$ and $0.31$ on $\mathrm{arcsinh}(\mathrm{SFR})$, with somewhat larger residuals for Mufasa and Simba (consistent with their smaller training samples).  sSFH reconstruction RMSE, averaged over galaxies and time bins, is $14.0$ on the area-normalised targets -- i.e.\ $\lesssim 3\%$ of the fixed target area of $500$ -- with the largest residuals concentrated at the high-SFR peaks where the normalisation amplifies dynamic range.  The latent representation itself is used in Appendix~\ref{app:dim_smooth} to probe the effective dimensionality of the sSFH manifold and the sensitivity of the classification signal to short-timescale variability.

\section{Dimensionality and smoothing ablation}\label{app:dim_smooth}

The interpretation developed in Section~\ref{sec:discussion} hinges on two claims about the sSFH data: that the simulation-of-origin signal lives in long-timescale morphology rather than in sub-Gyr burstiness, and that it is spread across enough independent directions of variation that a $2$-D UMAP embedding cannot hold all of them.  This appendix describes a series of ablation and diagnostic experiments on the multi-objective auto-encoder of Appendix~\ref{app:zehao}.

\subsection{Smoothing ablation\label{app:dim_smooth:smoothing}}

To test whether classification performance rides on short-timescale variability, we retrain the auto-encoder on sSFHs that have first been smoothed with a $1~\mathrm{Gyr}$ boxcar -- a $10$-bin uniform kernel on our $0.1~\mathrm{Gyr}$ grid -- applied either directly to the sSFH amplitudes (a \emph{linear} smoothing) or in $\mathrm{arcsinh}$-transformed space, where bursts are compressed before averaging and restored afterwards (an \emph{arcsinh} smoothing).  The kernel is applied before the area normalisation of Equation~\ref{eq:ae_norm}, so every galaxy's smoothed sSFH integrates to the same target area as its unsmoothed counterpart.  Remaining hyperparameters match the Appendix~\ref{app:zehao} configuration. We run the ablation on the unweighted variant to keep the smoothing signal disentangled from the inverse-frequency class weights (although we find the results remain unchanged in the weighted case, as with Appendix \ref{app:zehao}).  Table~\ref{tab:ae_smooth} collects the results.

\begin{table}[t]
\centering
\caption{Effect of pre-training $1~\mathrm{Gyr}$ boxcar smoothing on auto-encoder performance.  Classification accuracy is reported for the simulation-of-origin head; mass RMSE is reported for the $\log_{10}(\SM/\MSun)$ regression target.  All numbers are out-of-sample under $5$-fold cross-validation on the unweighted variant.}
\label{tab:ae_smooth}
\begin{tabular}{lccc}
\hline
Pre-training smoothing & Accuracy & Mass RMSE (dex) \\
\hline
None (baseline) & $95.7\%$ & $0.238$ \\
Linear, $1~\mathrm{Gyr}$ boxcar & $95.2\%$ & $0.226$ \\
$\mathrm{arcsinh}$, $1~\mathrm{Gyr}$ boxcar & $95.4\%$ & $0.196$ \\
\hline
\end{tabular}
\end{table}

Removing sub-Gyr structure degrades classification by only $0.3$--$0.5$ percentage points -- well within the fold-to-fold scatter of the baseline.  A kernel this wide would be expected to erase any genuinely bursty signal, so the near-invariance of the classifier implies that the features distinguishing simulations live on timescales $\gtrsim 1~\mathrm{Gyr}$.  The mass regression head, meanwhile, \emph{improves} under smoothing, most clearly in the $\mathrm{arcsinh}$ variant: sub-Gyr variance behaves as noise on the integrated $z=0$ properties that the regression is trying to recover, and removing it sharpens the recovery.

\subsection{UMAP on smoothed sSFHs\label{app:dim_smooth:umap}}

\begin{figure*}[hbtp!]
    \centering
    \includegraphics[width=\textwidth]{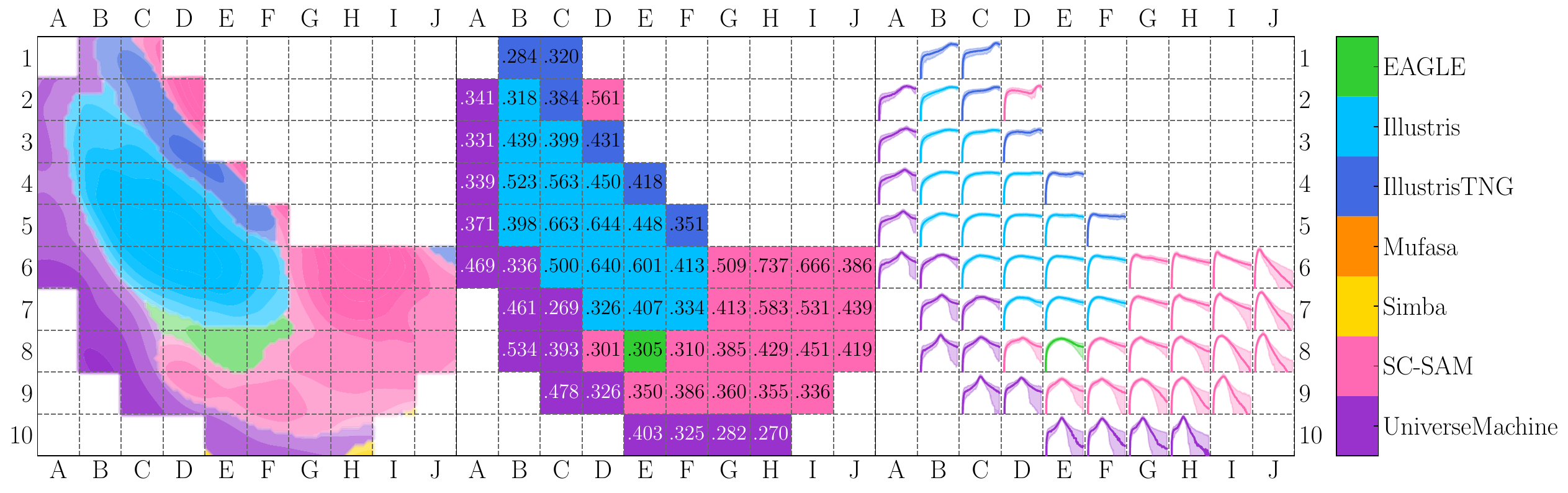} 
    \caption{SOPs on the UMAP of the 1-Gyr-boxcar-smoothed sSFHs following the exact layout of Figure~\ref{fig:sop} (\textbf{left}: each mesh point coloured by the dominant simulation with filled contours tracing the dominant SOP value; \textbf{centre}: cell-averaged dominant SOP on the battleship grid (value printed); \textbf{right}: cell-median sSFH coloured by the dominant simulation).  The macroscopic structure of the taco and each simulation's relative dominant neighbourhood is preserved under smoothing. In particular, the Illustris-dominated cluster at C6 (Figure~\ref{fig:UMAP}) remains in the same relative place (now $\sim$C5 in this figure), confirming that the feature reflects a long-timescale morphological signature rather than sub-Gyr burstiness.}
    \label{fig:umap_smoothed}
\end{figure*}

The same $1~\mathrm{Gyr}$ boxcar is applied to the sSFHs before recomputing the UMAP embedding with the identical hyperparameter and random-seed configuration used in the main text (Section~\ref{sec:methods}).  Figure~\ref{fig:umap_smoothed} shows the resulting embedding in the style of the SOP map of Figure~\ref{fig:sop} for direct comparison. 

The two embeddings share the same macroscopic layout, and the Illustris-dominated C6 cluster ($\sim$C5 cluster with the smoothed embedding) retains both its position and its dominant code composition.  Whatever distinguishes the C6 Illustris-dominated cells (Section~\ref{sec:discussion}) therefore must depend on features that survive $1~\mathrm{Gyr}$ smoothing.

\subsection{Classifier performance by input representation\label{app:dim_smooth:classifiers}}

The auto-encoder's high classification accuracy could reflect either the information content of its $16$-dimensional latent space or the expressive power of the classification head reading it.  We disentangle the two by training independent classifiers of increasing capacity -- a linear logistic (i.e. softmax) regression and a two-hidden-layer MLP ($100$--$100$, ReLU) -- on three representations of the same $63{,}083$-galaxy sample: the raw $136$-D sSFH, the auto-encoder's $16$-D latent, and the $2$-D UMAP embedding used throughout the main text.  Both classifiers share the $5$-fold cross-validation protocol of Appendix~\ref{app:zehao}.  Table~\ref{tab:classifier_reps} reports the results, with the auto-encoder's built-in head included for reference.

\begin{table}[t]
\centering
\caption{Simulation-of-origin classification accuracy on three input representations of the sSFH sample, for a logistic-regression baseline (``LogReg''), a two-hidden-layer MLP ($100$--$100$, ReLU), and -- for the $16$-D auto-encoder latent only -- the built-in classification head of the multi-objective auto-encoder (``AE head''; Appendix~\ref{app:zehao}).  All numbers are out-of-sample under $5$-fold cross-validation on the unweighted variant.}
\label{tab:classifier_reps}
\begin{tabular}{lccc}
\hline
Input representation & LogReg & MLP & AE head \\
\hline
UMAP, $2$-D & $43.7\%$ & $49.8\%$ & --- \\
AE latent, $16$-D & $54.9\%$ & $94.3\%$ & $95.7\%$ \\
Raw sSFH, $136$-D & $66.3\%$ & $89.2\%$ & --- \\
\hline
\end{tabular}
\end{table}

The MLP on the $2$-D UMAP coordinates caps at $49.8\%$ regardless of the capacity stacked on top of it.  On the $16$-D auto-encoder latent, the same MLP reaches $94.3\%$ -- within $1.4$ percentage points of the auto-encoder's built-in head -- and it exceeds the MLP on the full $136$-D raw sSFH ($89.2\%$).  The latent is therefore not merely dimensionally richer than UMAP's $2$-D embedding; it has distilled the classification-relevant directions of variation into a form more accessible than the raw input itself.  UMAP, by contrast, apparently is bottlenecked by its ambient dimensionality rather than by the classifier reading its coordinates.

\subsection{Intrinsic dimensionality of the representations\label{app:dim_smooth:twonn}}

To quantify how much of the sSFH manifold each representation retains, we estimate its intrinsic dimensionality using the TwoNN nearest-neighbour estimator of \citet{Facco_2017}.  TwoNN exploits the scale-free distribution of the ratio of second- to first-nearest-neighbour distances: under local uniformity, this ratio has a cumulative distribution whose power-law tail exponent equals the intrinsic dimensionality of the data manifold.  The estimator is applied to each representation of the $\sim$60k galaxy sample, with no subsampling.

\begin{table}[t]
\centering
\caption{Intrinsic dimensionality estimates from the TwoNN estimator of \citet{Facco_2017} for the three representations of the sSFH sample used in Table~\ref{tab:classifier_reps}, together with the auto-encoder latent trained on $1~\mathrm{Gyr}$-smoothed sSFHs for comparison.  ``Ambient'' is the dimensionality of the representation as stored; ``Intrinsic'' is the TwoNN estimate of the effective manifold dimension.}
\label{tab:intrinsic_dim}
\begin{tabular}{lcc}
\hline
Representation & Ambient & Intrinsic \\
\hline
UMAP, $2$-D & $2$ & $2.0$ \\
AE latent, $16$-D & $16$ & $9.0$ \\
AE latent, $1~\mathrm{Gyr}$-smoothed, $16$-D & $16$ & $8.3$ \\
Raw sSFH, $136$-D & $136$ & $23.8$ \\
\hline
\end{tabular}
\end{table}

The raw sSFH is effectively $\sim\! 24$-dimensional -- an order of magnitude below its $136$-D ambient storage, as expected for a smoothly correlated time series -- and the auto-encoder compresses this further into $\sim\! 9$ effective dimensions while sharpening rather than eroding the classification signal of Table~\ref{tab:classifier_reps}: its $16$-D latent is only partly utilised, and the learned representation is intrinsically lower-dimensional than it looks.  The $2$-D UMAP embedding, by contrast, saturates at its ambient dimensionality, which is what sets the hard ceiling on its classifier performance in Table~\ref{tab:classifier_reps}.  Smoothing trims the auto-encoder latent's intrinsic dimensionality from $9.0$ to $8.3$ -- a small but non-zero drop, consistent with the mild classification degradation under smoothing and suggesting that sub-Gyr variability contributes of order one effective dimension of structure to the latent, even though it is not where the dominant code-level signal lives.

The three tables help support the interpretation advanced in Section~\ref{sec:discussion} that the gap between UMAP-based and auto-encoder-based classification accuracy is dimensional, not ``textural'' (i.e. dominated by burstiness/short-timescale features). This means that simulation-of-origin information is distributed across several independent directions of Gyr-scale sSFH morphology, and a $2$-D embedding cannot hold all of them at once.  The auto-encoder's advantage over UMAP is therefore only representational capacity for long-timescale shape, not sensitivity to sub-Gyr burstiness.

\section{Application to the study of Milky Way analogues}\label{app:MWAs}
A simple application of our methods is to the study of galaxies similar to our own, \textit{Milky Way analogues (MWAs)}, to answer questions such as: Do the MWAs in our galaxy sample have a consistent sSFH profile? Are their shapes commonly produced by all simulations in our sample, or unique to one?  

As an example, we define ``MWAs" following \citet{Savelli2025} as galaxies with similar present-day stellar mass (SM) and SFR to the Milky Way, and quantify a galaxy's ``Milky Way-ness" using their population weights method, which we will briefly summarize here.  First apply a $\chi^2$ distribution to the galaxy sample:
\begin{equation}
    \chi_i^2=\left(\frac{M_{\star,i}-\rho_\textrm{SM}}{\Delta_\textrm{SM}}\right)^2+\left(\frac{\dot{M}_{\star,i}-\rho_\textrm{SFR}}{\Delta_\textrm{SFR}}\right)^2,
    \label{eq:chi2}
\end{equation}
where the subscript $i\in[0,n]$ denotes the $i^\textrm{th}$ galaxy in the sample of $n$ total galaxies, $M_{\star,i}(z)$ and $\dot{M}_{\star,i}(z)$ are respectively the SM and SFR of all galaxies in the sample, and $\rho_p$, $\Delta_p$ are respectively the value and uncertainty reported by \citet{Licquia_2015} for each parameter $p\in\{\textrm{SM,\ SFR}\}$, given as
\begin{align}
    \rho_\textrm{SM} \pm \Delta_ \textrm{SM}&=6.08\pm1.14\times10^{10}\ \mathrm{M_\odot},\label{eq:LNB-SM}\\
    \rho_ \textrm{SFR} \pm \Delta_ \textrm{SFR} &= 1.65 \pm 0.19\ \MSun\ \mathrm{yr}^{-1}.
    \label{eq:LNB-SFR}
\end{align}
The MWAs are then identified by the weights, $W_{i}$, from the $\chi^2$ distribution in Equation~\ref{eq:chi2}:
\begin{equation}
    \left\{W_{i}\right\}=\left\{\exp\left(-\frac{\chi_i^2}{2}\right)\right\}.
    \label{eq:W0}
\end{equation}
This results in a slight skew towards low-SM, low-SFR galaxies, which is then offset via dividing through by the pdf from a Gaussian KDE over SM and SFR
\begin{equation}
    \gamma=\frac{\left\{W_{i}\right\}}{\textrm{pdf}}=\left\{\exp\left(-\frac{\chi_i^2}{2}\right)/\textrm{pdf}\right\},
    \label{eq:W}
\end{equation}
culminating in the final set of weights used as the Milky Way-ness metric, denoted $\gamma$.  An effective sample size of MWAs can also be calculated from these weights as
\begin{equation}
    N=\frac{\left(\sum_{i=0}^n W_i\right)^2}{\sum_{i=0}^n\left(W_i^2\right)}.
    \label{eq:ess}
\end{equation}

Applying the $\gamma$ weights to our galaxy sample results in Figure~\ref{fig:MWA}, which follows the same layout as Figure~\ref{fig:kde}.  The left panel shows all galaxies in the UMAP projection space coloured by $\gamma$, with 0 being least similar to the Milky Way and 1 being most similar, and the centre and right panels show the cell-averaged effective sample size of MWAs.  Most MWAs live in the upper half of the taco, with the strongest clustering about the perimeter around cells D3, E3, and F3.  The associated sSFH display general shapes of early assembly with moderate-to-steep quenching, in line with the results from ~\citet{Savelli2025}.  Interestingly, the region most-populated with MWAs lies along the commonality strip in Figure~\ref{fig:shannon}, indicating most simulations in our sample are able to produce MWAs.

\bibliography{Paper}{}
\bibliographystyle{aasjournalv7}

\end{CJK*}
\end{document}